\documentclass[aps,prf,onecolumn,citesort,superscriptaddress]{revtex4-2}
\usepackage{graphicx}
\usepackage{newtxtext}
\usepackage{newtxmath}
\usepackage{scalerel}
\usepackage{bm}
\usepackage{natbib}
\usepackage[font=small,labelfont=bf,format=plain,justification=justified]{caption}
\usepackage{hyperref}
\hypersetup{
    colorlinks = true,
    urlcolor   = blue,
    citecolor  = blue,
}

\newcommand{\RomanNumeralCaps}[1]
\linenumbers

\def\Saad#1{{\color{black}#1}}

\begin{document}

\title{Impact of Particle Injection Velocity on the Stability of the Particulate Rayleigh-Bénard System}

\author{Saad Raza}
\email{saad.raza@univ-lille.fr}
\affiliation{Univ.~Lille, Unit\'e de M\'ecanique de Lille - J.~Boussinesq~(UML)~ULR~7512, F-59000 Lille, France}
\author{Romulo B. Freitas} 
\affiliation{Federal Center for Technological Education Celso Suckow da Fonseca, Nova Iguaçu-RJ 26041-271, Brazil}
\author{Leonardo S. B.  Alves}
\affiliation{PGMEC, Univ. Federal Fluminense, Rua Passo da Patria 156, Niteroi, Rio de Janeiro 24210-240,
Brazil}
\author{Enrico Calzavarini}
\affiliation{Univ.~Lille, Unit\'e de M\'ecanique de Lille - J.~Boussinesq~(UML)~ULR~7512, F-59000 Lille, France}
\author{Silvia C. Hirata}
\email{silvia.hirata@univ-lille.fr}
\affiliation{Univ.~Lille, Unit\'e de M\'ecanique de Lille - J.~Boussinesq~(UML)~ULR~7512, F-59000 Lille, France}
%\affiliation{Univ.~Lille, Unit\'e de M\'ecanique de Lille - J.~Boussinesq~(UML)~ULR~7512, F-59000 Lille, France}
%\affiliation{Federal Center for Technological Education Celso Suckow da Fonseca, Nova Iguaçu-RJ 26041-271, Brazil}
%\affiliation{PGMEC, Univ. Federal Fluminense, Rua Passo da Patria 156, Niteroi, Rio de Janeiro 24210-240,
%Brazil}

\date{\today}

\begin{abstract}
The linear stability of a thermally stratified fluid layer between horizontal walls, where \Saad{non-Brownian} thermal particles are continuously injected at one boundary and extracted at the other—a system known as particulate Rayleigh-Bénard (pRB)- is studied. For a fixed volumetric particle flux \Saad{and minimal thermal coupling}, reducing the injection velocity stabilizes the system when heavy particles are introduced from above, but destabilizes it when light particles are injected from below. 
For very light particles (bubbles), low injection velocities can shift the onset of convection to negative Rayleigh numbers, i.e. heating from above. 
Particles accumulate non-uniformly near the extraction wall and in regions of strong vertical flow, aligning with either wall-impinging or wall-detaching zones depending on whether injection is at sub- or super-terminal velocity. \Saad{The increase of the volumetric particle flux always enhances these effects.}
\end{abstract}

\maketitle

\section{Introduction}
The stability of a quiescent fluid layer under the influence of a settling dispersed phase composed of particles, drops, or bubbles, is a fluid dynamic problem of remarkable richness and complexity \citep{Guazzelli2011,mudde2005}. Even under highly idealized physical conditions, the problem statement involves a large number of physical parameters required to specify the material properties of both the fluid and the suspension, as well as all their boundary conditions. This is even more true when the couplings between the fluid and particles is not only mechanical but also energetic, involving e.g. temperature, phase-changes (melting, condensation, evaporation). 
In the case where the particles are small, very numerous, and highly diluted, an Eulerian two-fluid approach can be adopted to identify the parametric conditions under which they can destabilize the fluid, leading to large-scale advective motion \citep{Saffman_1962}.

Recently, \citet{prakhar2021linear} proposed studying a system in which a fluid layer is confined between two horizontal plates maintained at different temperatures, using heating from below to create an unstable density stratification (Rayleigh-B\'enard configuration). In this system, particles heavier than the fluid are continuously introduced from the top wall, at their terminal velocity and prescribed temperature, and removed from the bottom one. This system, named particulate Rayleigh-B\'enard (pRB), is shown to be more stable than the particle-free system. In other words, the Rayleigh number of the system must be larger than the $Ra_c \simeq 1708$ to trigger large-scale fluid motion, where $Ra_c$ identifies the supercritical bifurcation point of a pure fluid system. Subsequently, \citet{razaPF2024} extended the pRB model to particles of arbitrary density, including particles lighter than the fluid, which are injected from the lower plate at fixed temperatures. Even in this case, the system with particles is more stable than the particle-free system, regardless of the fluid-to-particle relative mass density and the strength of the mechanical and/or thermal couplings between the dispersed and continuous phases.
The latter result appears to contradict the analysis of \citet{NakamuraPhysRevE.102.053102, Nakamura_Yoshikawa_Tasaka_Murai_2021}, who demonstrated that the injection of bubbles from the bottom into an isothermal fluid layer is linearly unstable. Differently from the previously mentioned studies, these authors had considered the possibility of injecting bubbles at sub-terminal velocities, which is more realistic with respect to experiments. This fact has an important technical consequence in the linear stability problem, it allows the particle concentration to develop spatial inhomogeneities, which are prevented when the particles are injected at terminal velocity. Nakamura \textit{et al.} observed that the variation of the injection velocity does not affect the system stability, unless the velocity is very close to the terminal rising velocity.

These issues raise the question of the role of particle injection velocity in the pRB system, and whether there exists specific combinations of injection velocities and particle densities allowing the system to be de/stabilized and, thus, controlled. They are addressed in the present study by improving the linear stability analysis of \cite{razaPF2024}, where the particle injection velocity was set equal to the terminal velocity. Doing so shows that the system can be either stabilized or destabilized by properly tuning the particle inlet velocity and flux, where heavy and light particles induce opposite trends. 

\section{The particulate Rayleigh-B\'enard model system}
Following  \cite{razaPF2024}, we adopt an Eulerian model to describe the dynamics of a macroscopic material particle suspension in the pRB setting. The particle volume concentration is assumed to be small throughout, allowing the fluid to be treated as incompressible and governed by the Navier-Stokes equations under the Boussinesq approximation for the velocity \(\textbf{u}(\bm{x},t)\) and temperature \(T(\bm{x},t)\) fields.  However, due to the total conservation of momentum and thermal energy, the particulate-phase exerts both mechanical and thermal feedback on the fluid. This phase is characterized by the individual material properties of the particles, including mass density \(\rho_p\), diameter \(d_p\), and specific heat capacity at constant pressure \(c_{Pp}\). Additionally, it is described by the volume concentration \(\alpha(\bm{x},t)\), velocity \(\textbf{w}(\bm{x},t)\), and temperature \(T_p(\bm{x},t)\) fields. The governing conservation equations for mass, momentum, and heat for both the fluid and particle phases are given as follows:  
\begin{eqnarray}
      0 &=&  \nabla \cdot \textbf{u}, \label{eq:mass-fluid}\\
d_t \alpha  &=& - \alpha (\nabla \cdot \textbf{w}),\label{eq:mass-particles}\\
 D_t\mathbf{u}  &=& \tfrac{-\nabla p}{\rho} +\nu\nabla^2 \mathbf{u}   + [1-\beta_T(T-T_r)]\mathbf{g} + \alpha \left[ \left(D_t\mathbf{u} - \mathbf{g}\right) + \tfrac{\rho_p}{\rho} \left(\mathbf{g} - d_t\mathbf{w}  \right) \right], \label{eq:mom-fluid}\\
d_t \mathbf{w} &=&  \beta D_t\mathbf{u}+\tau_p^{-1}(\mathbf{u}-\mathbf{w})+(1-\beta)\mathbf{g},\label{eq:mom-particle}\\
   D_t T   &=& \kappa  \nabla^2  T  + \alpha \left[ {D_t T} -  E\ \tau_{T}^{-1}(T - T_p)\right],\label{eq:temp-fluid}\\
     d_t T_p&=& \tau_{T}^{-1}(T-T_p).\label{eq:temp-particle}
\end{eqnarray}

A few additional observations are in order.   First, $
D_t() = \partial_t () + \textbf{u} \cdot \nabla(),
$  and 
$
d_t() = \partial_t() + \textbf{w} \cdot \nabla(),
$ 
denote respectively the the fluid-phase and particulate-phase material derivative operators, where $\nabla()$ is the operator containing the spatial derivative vector. Second, we account for three primary hydrodynamic forces acting on the particles: (i) the Stokes drag force, (ii) the fluid acceleration force with the added mass correction, and (iii) buoyancy. The drag force is parameterized by the viscous response time  
$
\tau_p = d_p^2/(12 \nu \beta),
$  
where \(\nu\) is the fluid kinematic viscosity. The intensity of the added mass force is modulated by the modified density ratio  
$
\beta = 3\rho/(\rho + 2\rho_p),
$  
where \(\rho\) denotes the fluid mass density.  
Third, we assume that the temperature within each particle remains uniform (lumped approximation), with its relaxation towards equilibrium characterized by the timescale  
$
\tau_T = d_p^2 E/(12 \kappa),
$  
where \(\kappa\) is the fluid thermal diffusivity and 
$
E = \rho_p c_{Pp}/(\rho c_P),
$  
with \(c_P\) representing the fluid specific heat capacity at constant pressure.  
Finally, the remaining constants include the fluid volumetric thermal expansion coefficient \(\beta_T\) at the reference temperature \(T_r\), the gravitational acceleration vector \(\textbf{g}\), and the pressure field \(p(\textbf{x},t)\).  

The domain considered here is three-dimensional, infinitely long in both horizontal coordinates which are vertically bounded by two parallel horizontal walls located at \( z = \pm H/2 \), with the unit vector \( \hat{\textbf{z}} \) pointing upwards. The fluid satisfies no-slip boundary conditions at both walls (\(\textbf{u} = 0\)), which are kept isothermal, {with a thermal gap $\Delta T$.} The bottom wall is the warmest, creating an unstable density stratification when \( \beta_T > 0 \).  
Particles are introduced from one of the boundaries at a constant volumetric flux and with a prescribed velocity \( \mathbf{w}^* \), expressed as a multiple of the reference terminal velocity, i.e.  
$
\mathbf{w}_T = (1-\beta)\tau_p \mathbf{g}.
$  
Particles with \( \beta < 1 \), hereafter referred to as ``heavy particles'', are injected from the top boundary, whereas those with \( \beta > 1 \), i.e. ``light particles'', are injected from the bottom one. The particle inlet temperature is also set to a fixed value, \( T_p^* \), to be specified later.  
Particle accumulation at the opposite boundary is neglected, i.e., they are assumed to be removed from the domain as soon as they reach the opposite wall. 
It is worth noting that the equations governing the particulate-phase are first-order in space, as they lack a dissipation term in the form of a Laplacian. Consequently, only an inlet particulate-phase boundary condition is required to determine their solution.  

\subsection{Dimensionless system}
In preparation for the upcoming stability study, we re-wrote the model in dimensionless form using its characteristic height, \(H\), conductive time scale, \(\mathcal{T} = t \kappa/H^2\), and fluid density, \(\rho\). Accordingly, we define the following dimensionless fields:
$$\mathbf{U}=\mathbf{u}\tfrac{H}{\kappa}, \quad {P = \tfrac{p H^2}{\rho \kappa^2 }},
\quad\Theta = \tfrac{T-T_r}{\Delta T},
 \quad \mathbf{W}=\mathbf{w}\tfrac{H}{\kappa}, \quad\Theta_p = \tfrac{T_{p}-T_r}{\Delta T},$$
which represent the dimensionless fluid velocity, pressure, and temperature, as well as the particulate velocity and temperature. Maintaining the same notation for the dimensionless material derivatives and differential operator, and explicitly incorporating the particle momentum feedback, the governing equations (\ref{eq:mass-fluid})–(\ref{eq:temp-particle}) become:
\begin{eqnarray}
0 &=& \nabla \cdot \textbf{U},\label{massfluidunperturb}\\
d_{\mathcal{T}} \alpha &=& - \alpha (\nabla \cdot \textbf{W}),\label{massparticleunperturb}\\
D_{\mathcal{T}}\mathbf{U} &=& -\nabla P + Pr \nabla^2 \mathbf{U} + Pr Ra \Theta \hat{\mathbf{Z}} \nonumber \\
&+& \tfrac{\alpha}{2} \left[ (\beta-1) \left( D_{\mathcal{T}}\mathbf{U}+ {Ga Pr^{2} \hat{\mathbf{Z}}}\right) - 12 Pr (3-\beta )\Phi^{-2}\left( \mathbf{U} - \mathbf{W} \right) \right],
    \label{momentumfluidunperturb}\\
d_\mathcal{T} \mathbf{W}
 &=&  \beta \left( D_{\mathcal{T}}\mathbf{U} + 12 Pr \ \Phi^{-2} \left(\mathbf{U} - \mathbf{W}\right) \right) - {(1 - \beta)} {Ga Pr^{2} \hat{\mathbf{Z}}},
\label{momentumparticleunperturb}\\
D_{\mathcal{T}}\Theta &=& {\nabla^2 \Theta} + \alpha\left[{D_{\mathcal{T}}\Theta}- 12\Phi^{-2}\left(\Theta - \Theta_p \right)\right],
\label{energyfluidunperturb}\\
d_{\mathcal{T}} \Theta_p &=& 12 (E \Phi^{2})^{-1}\left(\Theta - \Theta_p \right),
    \label{energyparticleunperturb}
\end{eqnarray}
with dimensionless boundary conditions: 
\begin{equation}
\mathbf{U} = 0,\quad  \Theta = 1 \quad \textrm{at} \quad  Z=- \tfrac{1}{2}, \quad \textrm{and}\quad  
\mathbf{U} = 0;\quad  \Theta = 0 \quad \textrm{at} \quad  Z= \tfrac{1}{2},
\label{eq:bc:fluid} 
\end{equation}
\begin{equation}
  \mathbf{W}  = \mathbf{W^*} =W^*\hat{\mathbf{Z}},\qquad\alpha = \mathcal{J}/||\mathbf{W^*}||, \quad  \Theta_p =\Theta_p^*  \quad \textrm{at} \quad  Z=Z^*,\label{eq:bc:particle} 
\end{equation}
where $Z^*$ denotes the location of the inlet horizontal wall. Note that $W^*$ can be either positive or negative, depending on whether the particle is lighter or heavier than the fluid, while the inlet flux $\mathcal{J}$ is defined as always positive. In the above equations we have introduced the following dimensionless characteristic parameters:
\begin{equation}
Ra=\frac{\beta_T \Delta T g H^3}{\nu \kappa}, \quad Pr = \frac{\nu }{\kappa}, \quad Ga = \frac{g H^3}{\nu^2}, \quad \Phi = \frac{d_p}{H},\label{eq:control-param}
\end{equation}
where \( Ra \) denotes the Rayleigh number, which quantifies the relative strength of thermally induced buoyancy against mechanical and thermal dissipation, \( Pr \) the Prandtl number, which characterizes the fluid phase diffusive material properties, and \( Ga \) the Galileo number, which represents the balance between gravitational and viscous forces. Although $Ga$ does not depend on the particle properties, it becomes a relevant control parameter whenever the coupling between the particle and the fluid is taken into account. In dimensionless units, the terminal velocity expression is $\mathbf{W}_T =\frac{1-\beta}{\beta}\frac{ \Phi^2}{12} Ga Pr \hat{\mathbf{Z}}$. 
Along with the modified fluid-to-particle density ratio, \( \beta \), the volumetric particle inlet flux $\mathcal{J}$, and the inlet velocity and temperatures $(W^*, \Theta_p^*)$, they define the full set of control  parameters. In total, the model is governed by nine parameters: three associated with the fluid phase $( Ra, Pr, Ga )$ and six with the particulate-phase $( \Phi, \beta, E, \mathcal{J} , W^*, \Theta_p^* )$. 

\section{Linear stability analysis}
\subsection{Conductive state}
  
\begin{figure}
        \centering
        \includegraphics[width=0.40\linewidth]{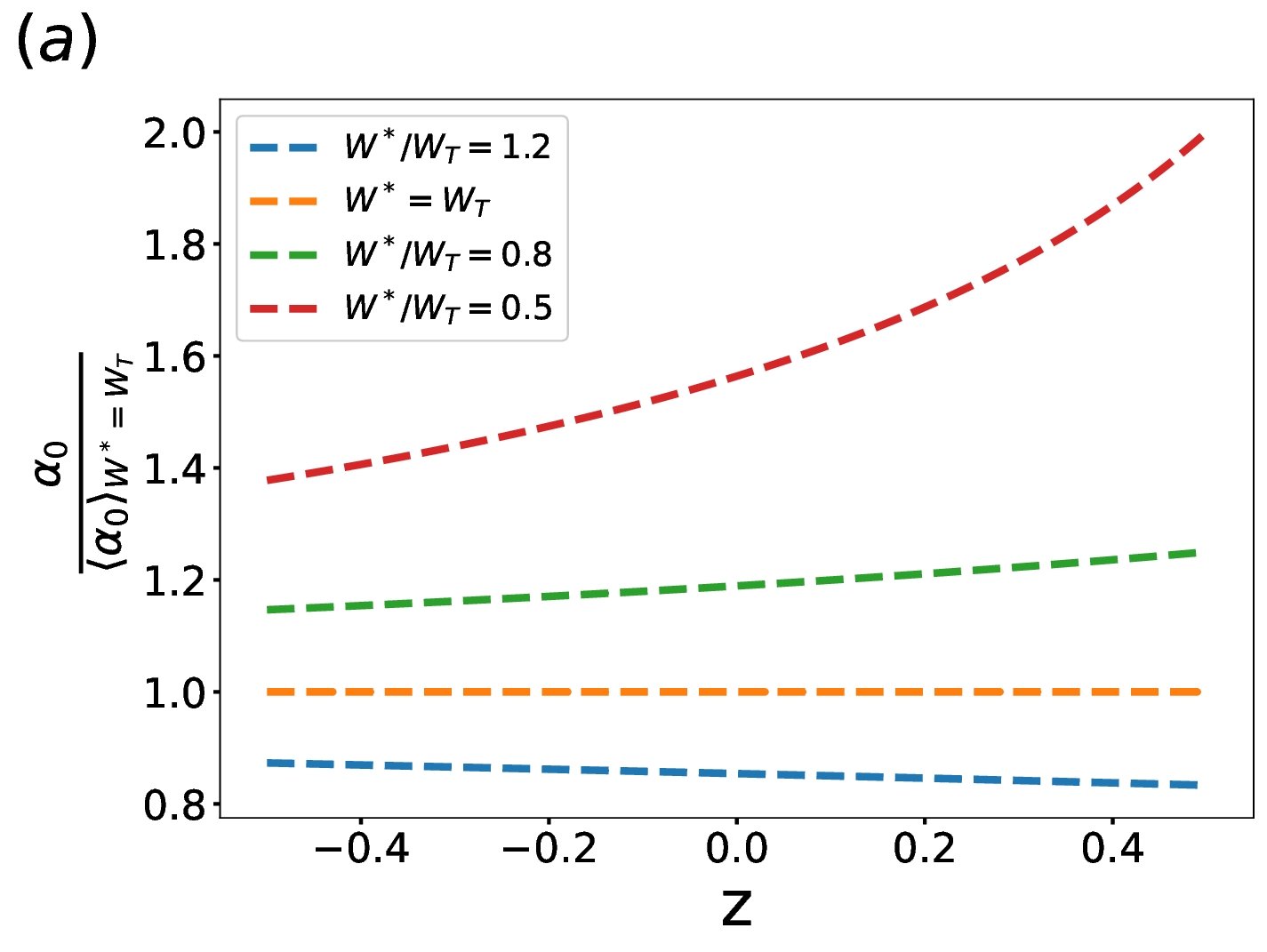}  
        \includegraphics[width=0.40\linewidth]{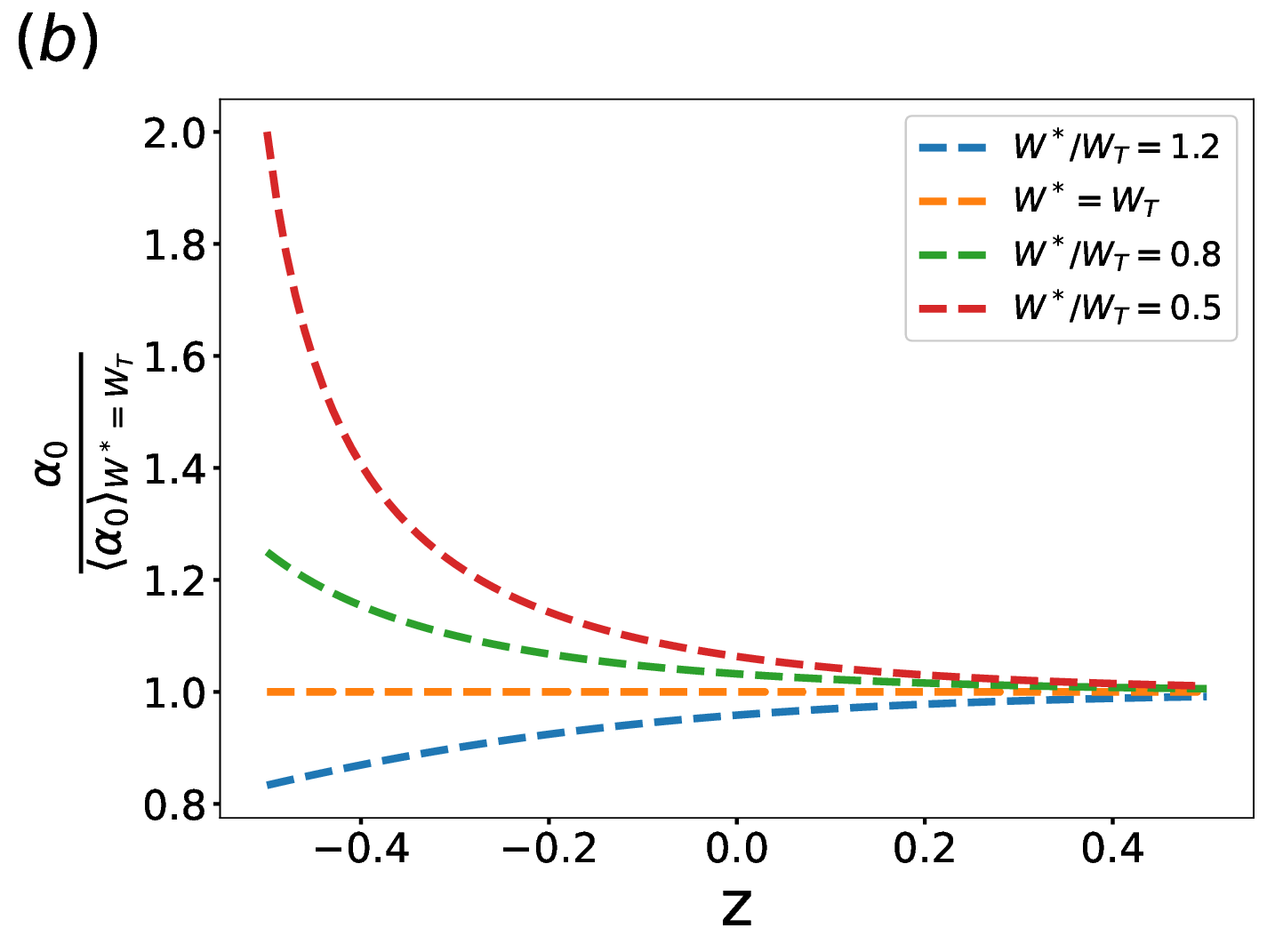} 
    \caption{Vertical dependence of the steady-state volume concentration field of the particulate-phase, $\alpha_0(Z)$, for different inlet velocities $W^*$ for (a) $\beta=0.5$ (heavy particles) and (b) $\beta=3$ (bubbles). Data are normalized by the steady-state concentration field corresponding to the case $W^*=W_T$.}
    \label{fig:base-state}
\end{figure}

In order to identify the onset of natural convection in the pRB system, its equilibrium solution must first be defined. The one chosen here is its steady-state with particles settling or rising in a quiescent and conductive fluid. Hence, one can impose $\mathbf{U}=0$, $\mathbf{W}=W_0(Z) \mathbf{\hat{Z}}$, $\alpha=\alpha_0(Z)$, $\Theta=\Theta_0(Z)$, and $\Theta_p=\Theta_{p0}(Z)$, and re-write equations (\ref{massfluidunperturb})-(\ref{energyparticleunperturb}) as:
\begin{equation}
    D(\alpha_0 W_0) = 0,
    \label{partice-mass-cons}
\end{equation}
\begin{equation}
    W_0 DW_0 = -{12 Pr \beta W_0}{\Phi^{-2}} - (1-\beta) {Ga Pr^{2} \hat{\mathbf{Z}}},
\end{equation}
\begin{equation}
     D^2 \Theta_0 - {12 \alpha_0}{\Phi^{-2}}(\Theta_0-\Theta_{p0}) = 0
     \label{eq16},
\end{equation}
\begin{equation}
 W_0 D\Theta_{p0} ={12}({E \Phi^2})^{-1} (\Theta_0-\Theta_{p0}),
\label{eq15}
\end{equation}
where $D$ represents the derivative with respect to $Z$. The steady-state governing equations (\ref{partice-mass-cons})-(\ref{eq15}), with their respective boundary conditions derived from (\ref{eq:bc:fluid}) and (\ref{eq:bc:particle}), are computed numerically. Particle volume fraction profiles are shown in figure \ref{fig:base-state}. It is possible to note that, in general, particle concentration varies with height, except for the special case where particles are injected at their terminal velocity ($W^*=W_T$), which is the case previously studied by \citet{prakhar2021linear} and \citet{razaPF2024}. 
When particles are introduced at a velocity lower (higher) than their terminal velocity, they accelerate (decelerate) until they reach it. This process leads to an accumulation (rarefaction) of particles near the injection wall, e.g. in the upper region for heavier particles and in the lower region for lighter ones. 
Finally, for the prescribed particle inflow considered here, reducing the injection velocity increases the volumetric concentration of incoming particles.
Hence, smaller inlet velocities lead to a higher amount of particles.

\subsection{Perturbation equations}

Having defined the equilibrium solution of the pRB system, the next step is to analyze its linear stability. This is done here by decomposing all variables into their respective steady-state and small amplitude perturbations, linearizing the resulting equations and assuming these perturbations be modeled as normal modes, such as
\begin{equation}
\xi'(X,Z,\mathcal{T})=\xi^n(Z) \exp(i k X+\lambda \mathcal{T}) + \textit{c.c.}\label{eq:normalmodes}
\end{equation}
where $\xi'=\{\mathbf{U}',\mathbf{W}', \alpha',\Theta',\Theta_p'\}$ is the vector of perturbed quantities, \textit{c.c.} means complex conjugate, and $\xi^n(Z)$ is the normal mode amplitude varying in the non-homogeneous direction $Z$. Furthermore, according to the temporal stability approach, $k$ is the real wave number, and $\lambda=\lambda_r+i\lambda_i$, where $\lambda_r$ is the temporal growth rate of the perturbation and $\lambda_i$ is its oscillation frequency.
Doing so, leads to their governing equations, namely
\begin{equation}
     \lambda \alpha^n  = -\big(\alpha_0 DW^n_{z} + W^n_{z}{D\alpha_0}\big)  - \alpha_0 \Dot{\iota} k W^n_{{x}} - \big(W_0 D\alpha^n  + \alpha^n {D W_0}\big),
     \label{eq30}
\end{equation}
\begin{equation}
\begin{aligned}
\lambda (D^2 -k^2) {U^n_z} &=  Pr (D^2 - k^2)^2 U^n_z - Pr Ra k^2 
\Theta^n + \frac{\alpha_0}{2} (\beta -1)  \lambda(D^2 -k^2) U_z^n \\
&+ {\frac{D\alpha_0 }{2} (\beta -1)\lambda D U_z^n - D\alpha_0{6 Pr (3 -\beta) }{\Phi^{-2}} (DU_z^n  + \dot\iota k W^n_x )}\\
&- \alpha_0 {6 Pr (3 -\beta) }{\Phi^{-2}}(\dot\iota k D W_x^n + (D^2-k^2) U_z^n +k^2 W_z^n)\\
&-  \Big[\frac{1}{2}(\beta -1){Ga Pr^{2}} + 6 Pr (3 -\beta) {W_0}{\Phi^{-2}}  \Big]k^2 \alpha^n,
\end{aligned}
\label{eq35}
\end{equation}

\begin{equation}
 \lambda {W_z}^n + W_0 {D {W_z}^n} +  {W_z}^n {DW_0} = \beta \lambda {U^n_z} + 12 Pr \beta \Phi^{-2}{({U_z}^n-{W_z}^n)},
 \label{eq37}
\end{equation}
\begin{equation}
 \lambda {W_x}^n + W_0 ({D{W_x}^n}) = \beta \lambda {U^n_x} + 12 Pr \beta  \Phi^{-2}{({U_x}^n-{W_x}^n)},
 \label{eq37}
\end{equation}
\begin{equation}
\begin{aligned}
       {(1-\alpha_0) } \left[\lambda {\Theta^n} + U^n_z D \Theta_0\right] &= \alpha^n \lambda \Theta_0 + (D^2 - k^2) \Theta^n   - 12 \alpha_0  \Phi^{-2}{(\Theta^n - \Theta_{p}^n)}\\
        &-  12 \alpha^n  \Phi^{-2}{(\Theta_0^n - \Theta_{p0}^n)},
\end{aligned}
 \label{eq38}
\end{equation}
\begin{equation}
     \lambda  \Theta^n_p + W_0 D\Theta^n_p + {W_z}^n  D \Theta_{p0}^n = {12}({E}{\Phi^2})^{-1}{(\Theta^n - \Theta^n_p)}.
     \label{eq39}
\end{equation}
obtained by taking the double curl of the fluid momentum equation to eliminate the pressure, and using the incompressibility relation in order to eliminate the $x-$component of the velocity. Their boundary conditions, obtained by similar means from (\ref{eq:bc:fluid}) and (\ref{eq:bc:particle}), are given by
\begin{eqnarray}
{U_z^n} = {DU_z^n} = 0, \quad  \Theta^n = 0  \quad   \text{at} \quad Z= \pm{1}/{2}\\
\alpha^n=0, \quad {\bm{W^n}} = 0,  \quad  
\Theta^n_{p} = 0\quad   \text{at} \quad Z = Z^*
\end{eqnarray}

Equations (\ref{eq30})-(\ref{eq39}) are solved numerically by employing the shooting method, where the critical conditions ($Ra_c, \lambda_c, k_c$) are determined from the sensitivity of $Ra$ with respect to $k$ (see \citet{Alvesetal} for details). In order to verify the accuracy of these results, a matrix-forming approach is also employed. It adopts a fourth-order finite-difference discretization to transform the differential system of equations into a generalized algebraic eigenvalue problem. The latter is solved numerically by the Arnoldi method with a shift-and-invert spectral transformation (see \cite{Souzaetal} for details). Although not shown here, the excellent agreement between both approaches verifies the accuracy of the numerical results presented in the next section.

\section{Results}

Since the present study focuses on the influence of particle inlet velocity on the model stability, the following parameters are kept constant throughout the analysis: $Ga = 12\times10^{12}$, $Pr=5$, $E = 5\times10^{-3}$, $\Phi = 10^{-2}$. Heavy particles are injected from the top with the cold wall temperature $\Theta_p^* = 0$, while light particles are injected from the bottom with the hot wall temperature $\Theta_p^* = 1$. Furthermore, the inlet flux was \textcolor{black}{varied with respect to a reference value} $\mathcal{J} = \mathcal{J}_0 = 533.3$.  Finally, the thermal feedback of the particles to the fluid is assumed negligible in the present work since $E\ll 1$.

\subsection{Time-asymptotic analysis}

\begin{figure}
    \centering
        \includegraphics[width=0.43\linewidth]{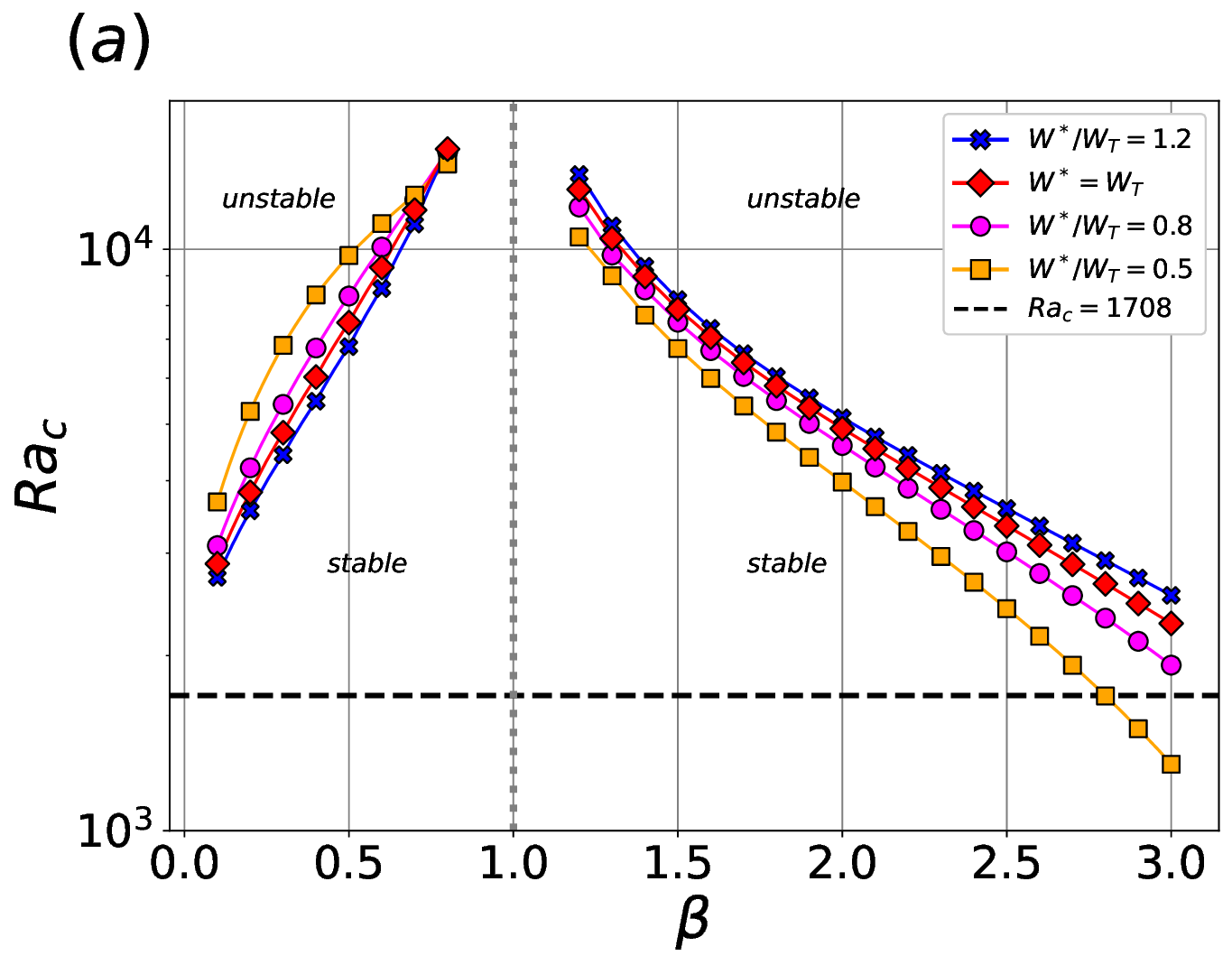}  
        \includegraphics[width=0.45\linewidth]{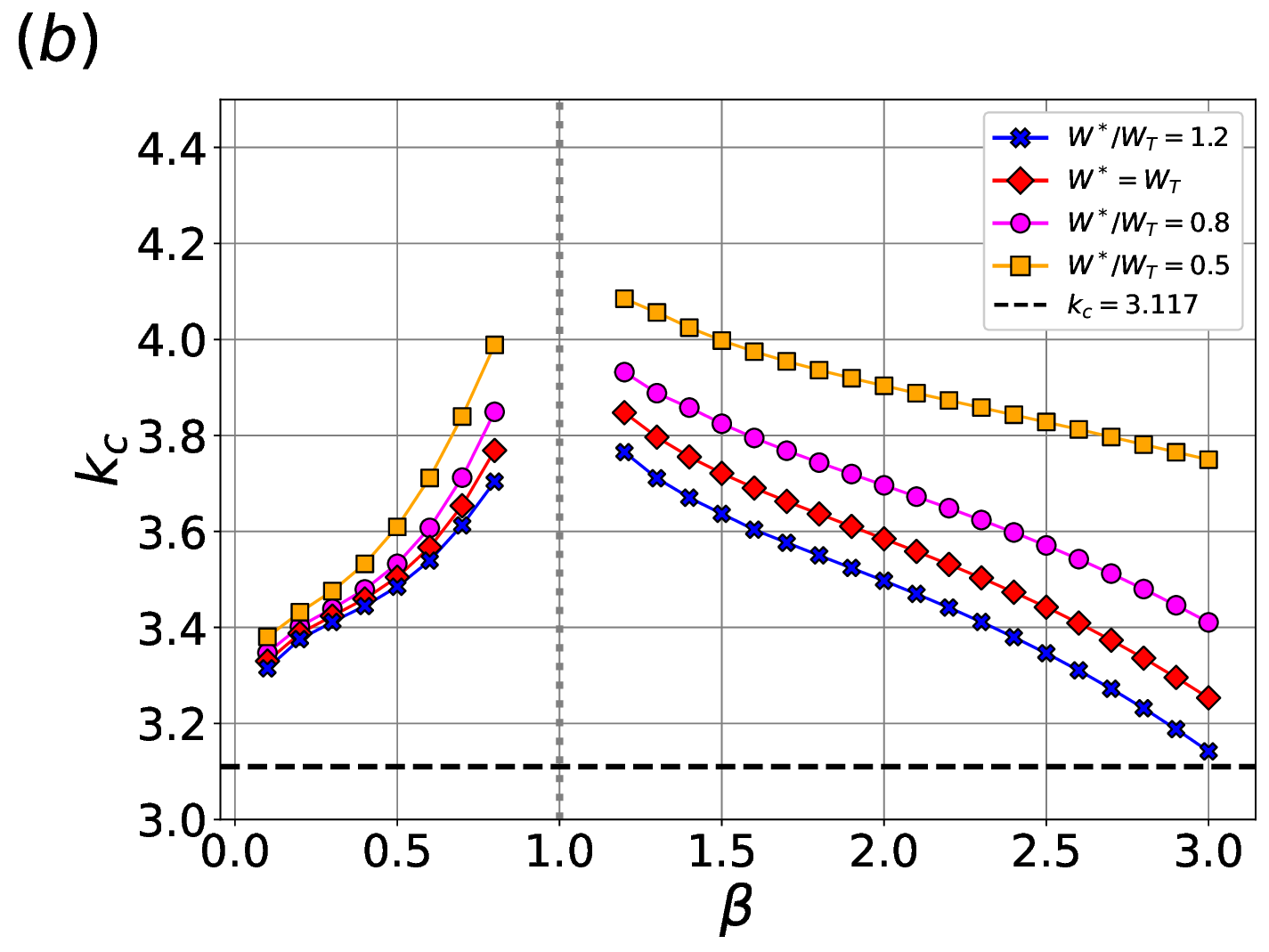}
    \caption{Variation with the density ratio $\beta$ of the critical Rayleigh number $Ra_c$ (a) and  wave vector $k_c$ (b).
    Calculations for various inlet velocities $W^*$ are shown, from sub-terminal $<W_T$ to super-terminal $>W_T$. For comparison the values $Ra_c,k_c$ corresponding to the single-phase RB system, are drawn as horizontal lines. \textcolor{black}{The particle inlet flux is $\mathcal{J} = \mathcal{J}_0$}.}
    \label{fig:Rac_beta}
\end{figure}
The first thing to notice is that the linear onset of instability is characterized by a pitchfork bifurcation in the parameter space explored here. In other words, the oscillatory frequency $\lambda_i$ is always zero.
Second, the effects of the modified density ratio on the critical thresholds under different particle inlet velocity are presented in figure \ref{fig:Rac_beta}. For heavy particles sedimenting from the top ($\beta<1$), an increase in the inlet velocity leads to a decrease in the critical Rayleigh number, thereby destabilizing the system. Conversely, for light particles injected from the bottom ($\beta>1$), the trend is reversed, with higher inlet velocities promoting stabilization. The most unstable cases occur for sub-terminal particle velocities. Notably, for $\beta > 2.8$ and $W^*/W_T=0.5$, the critical Rayleigh number for the pRB modal falls below that of the single-phase RB system, indicating that a weakly injected particulate-phase can promote instability in the case of light particles and bubbles. It should be noted that the case of neutrally buoyant particles ($\beta=1$) is singular, since one would have to inject an infinite amount of particles to keep the volumetric particulate flux constant. 
As illustrated in figure \ref{fig:Rac_beta}(b), the critical wave number slightly decreases with increasing inlet particle velocity for both heavy and light particles, though this decrease is more pronounced in the latter case.

\begin{figure}
    \centering
    \includegraphics[width=0.40\linewidth]{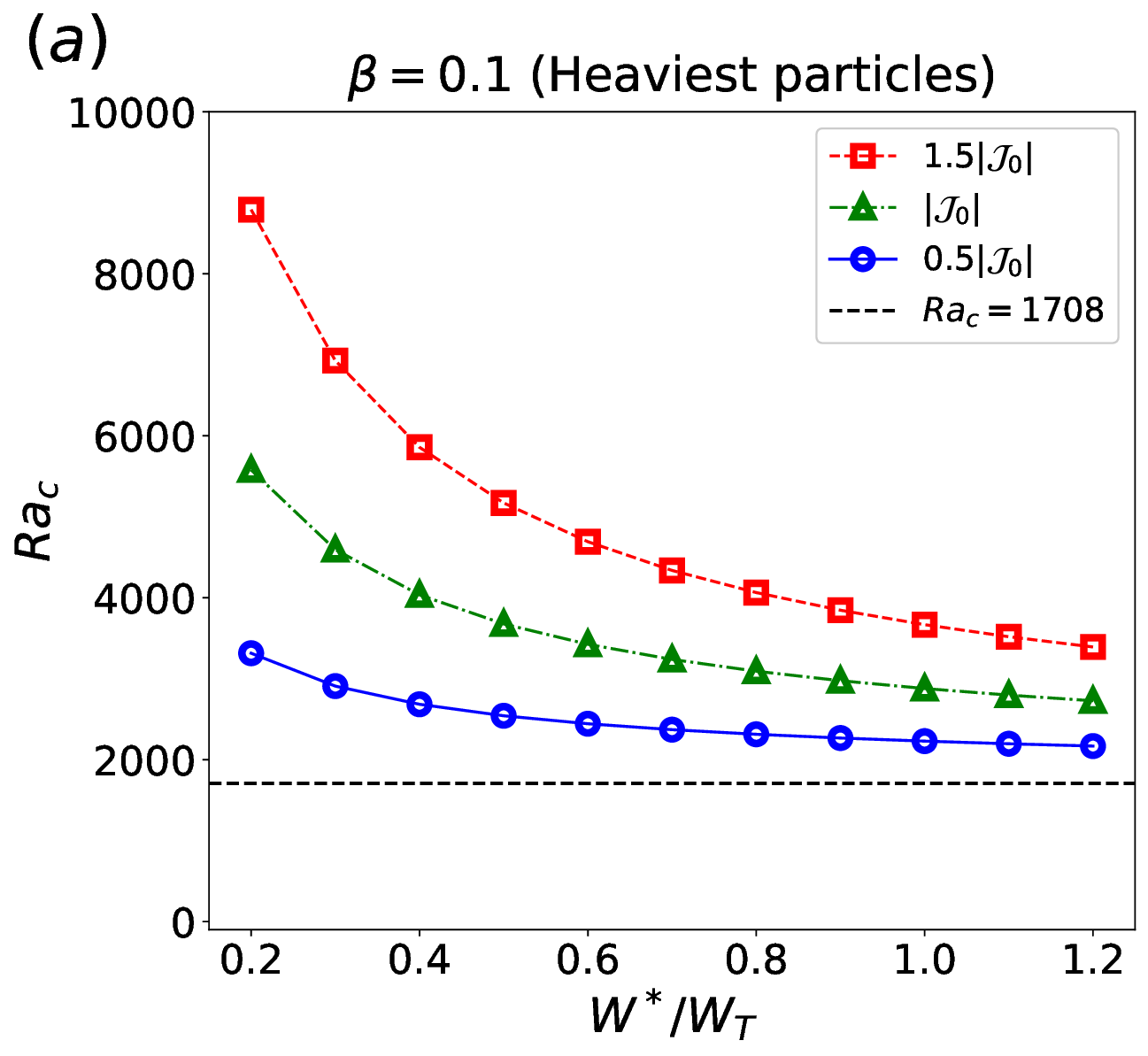} 
    \includegraphics[width=0.40\linewidth]{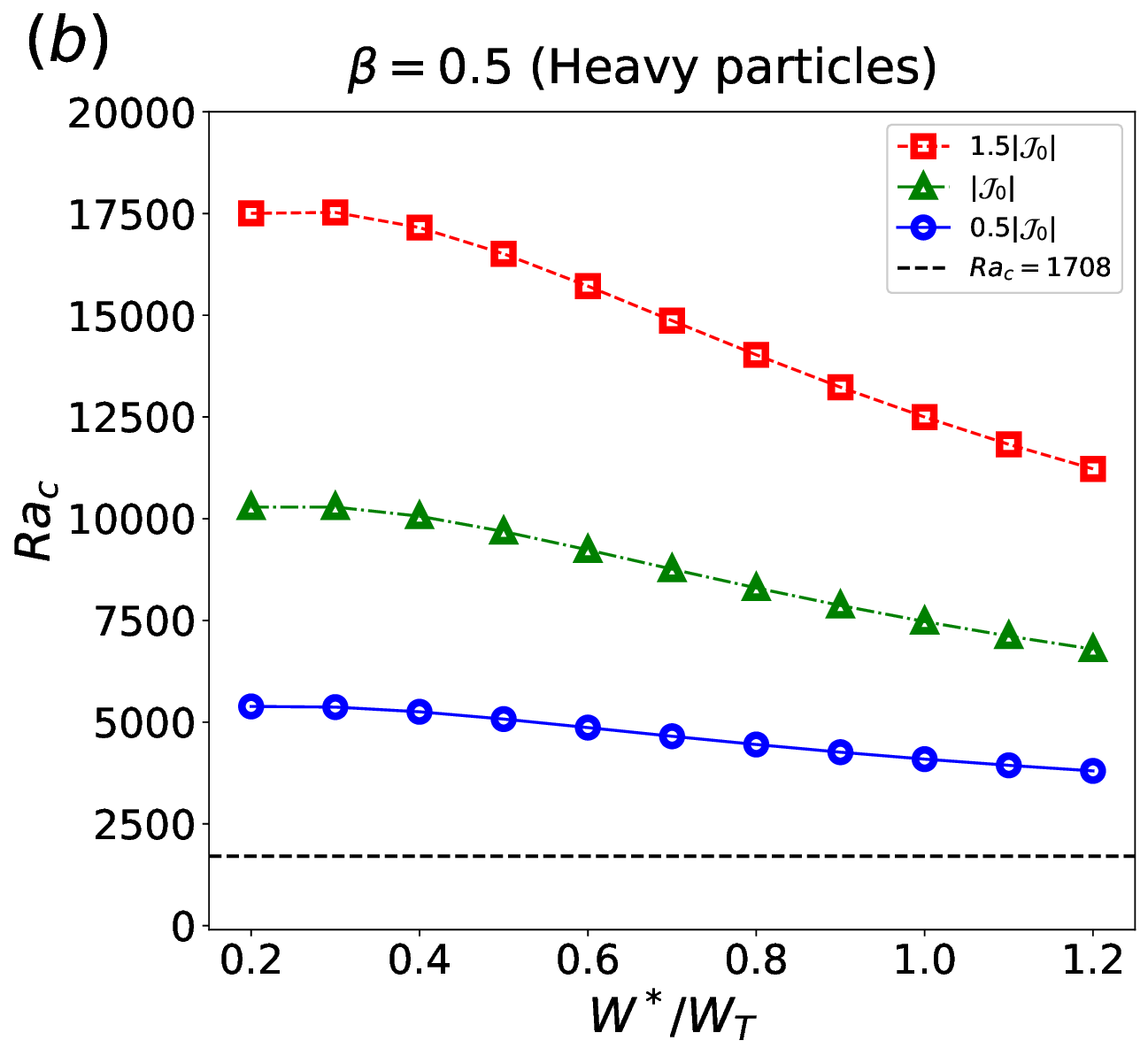} \\[1ex]
    \includegraphics[width=0.40\linewidth]{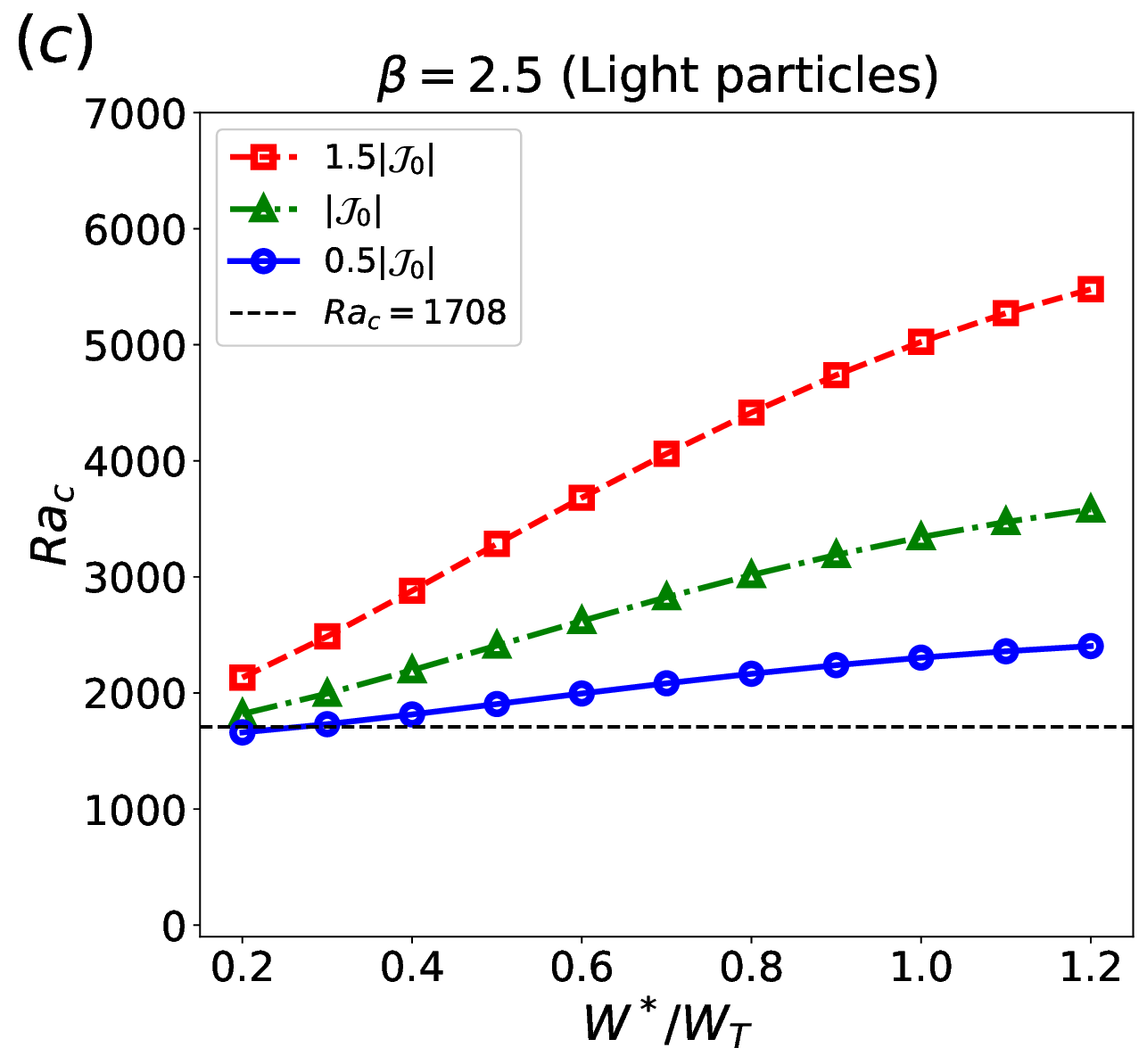} 
    \includegraphics[width=0.40\linewidth]{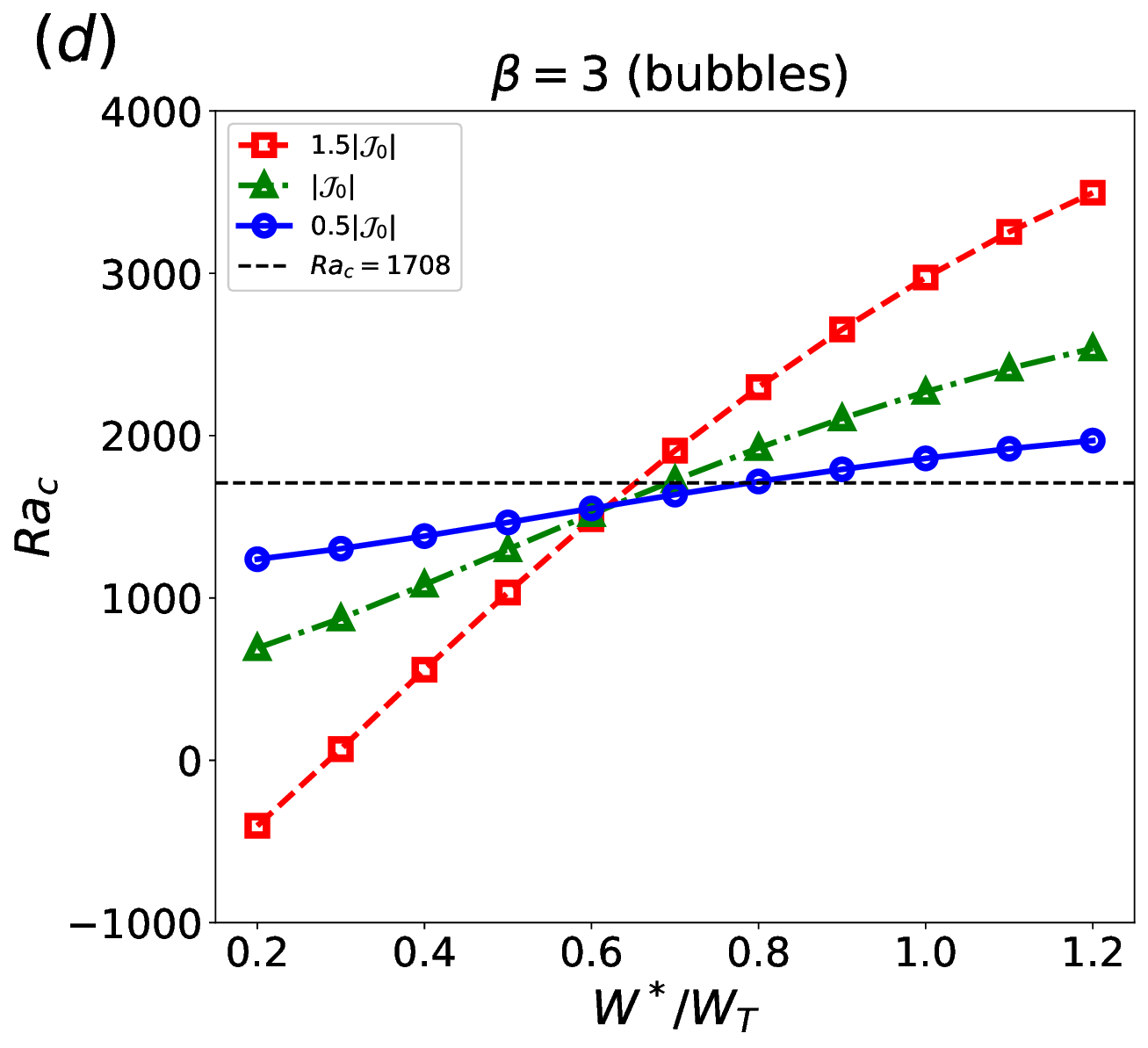} 
    \caption{Critical Rayleigh number as a function of the inlet velocity for three different particle fluxes. The horizontal dashed lines correspond to the single-phase RB threshold.}
    \label{fig:all_figures-flux}
\end{figure}

Figures \ref{fig:all_figures-flux}(a-d) show the effect of the particle inlet velocity on the critical thresholds under different inlet volumetric flux intensities for selected values of the modified density ratio. For the case of bubbles ($\beta=3$), increasing the inlet flux is destabilizing for low inlet velocities but stabilizing otherwise, since all three curves intersect at $W^*/W_T\sim0.6$. A similar trend is observed for all light particle cases, though the value of $W^*/W_T$ where the intersection occurs decreases as $\beta$ decreases. For $\beta \lesssim 2.5$, no intersection is observed and the increase in the particle flux always plays a stabilizing role. It is also worth pointing out that, for $\beta=3$ and $\mathcal{J}=1.5 \mathcal{J}_0$, the critical Rayleigh number becomes negative for low enough values of $W^*/W_T$. This means that bubbles can trigger convective instabilities even when the pRB model is heated from above. As discussed by \cite{NakamuraPhysRevE.102.053102}, this is due to a potentially unstable density stratification within the liquid-gas mixture near the bottom wall. As the bubble velocity increases to its terminal velocity, the volume fraction in this region decreases from $\mathcal{J} /W^*$ to $\mathcal{J} /W_T$, increasing the possibility of a locally unstable density gradient. 

\begin{figure}
\setkeys{Gin}{width=1\linewidth}
\begin{minipage}[t]{0.24\textwidth}
\includegraphics[width=\linewidth]{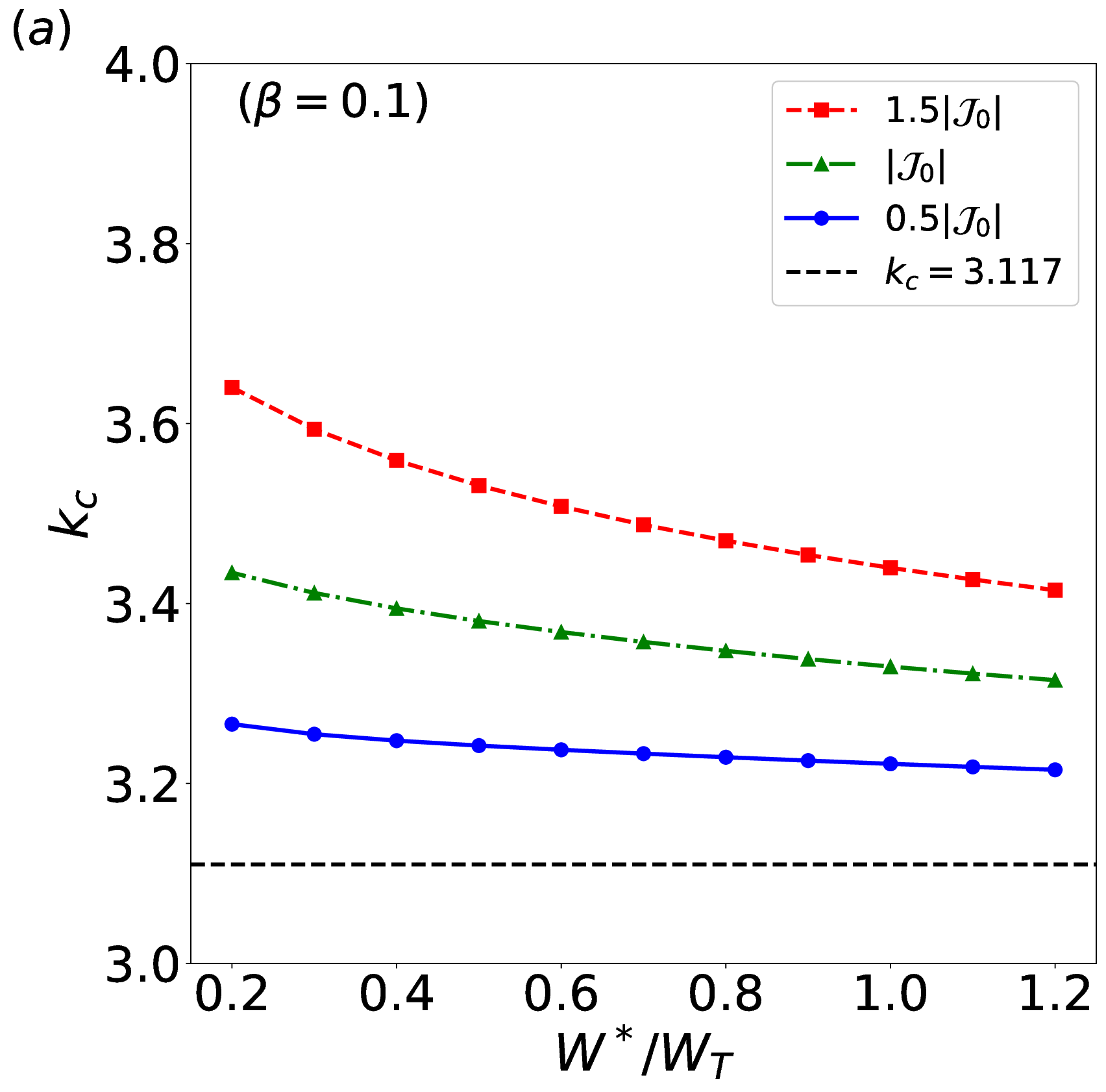}
  \label{fig:fig1}
\end{minipage}\hfill
\begin{minipage}[t]{0.24\textwidth}
\includegraphics[width=\linewidth]{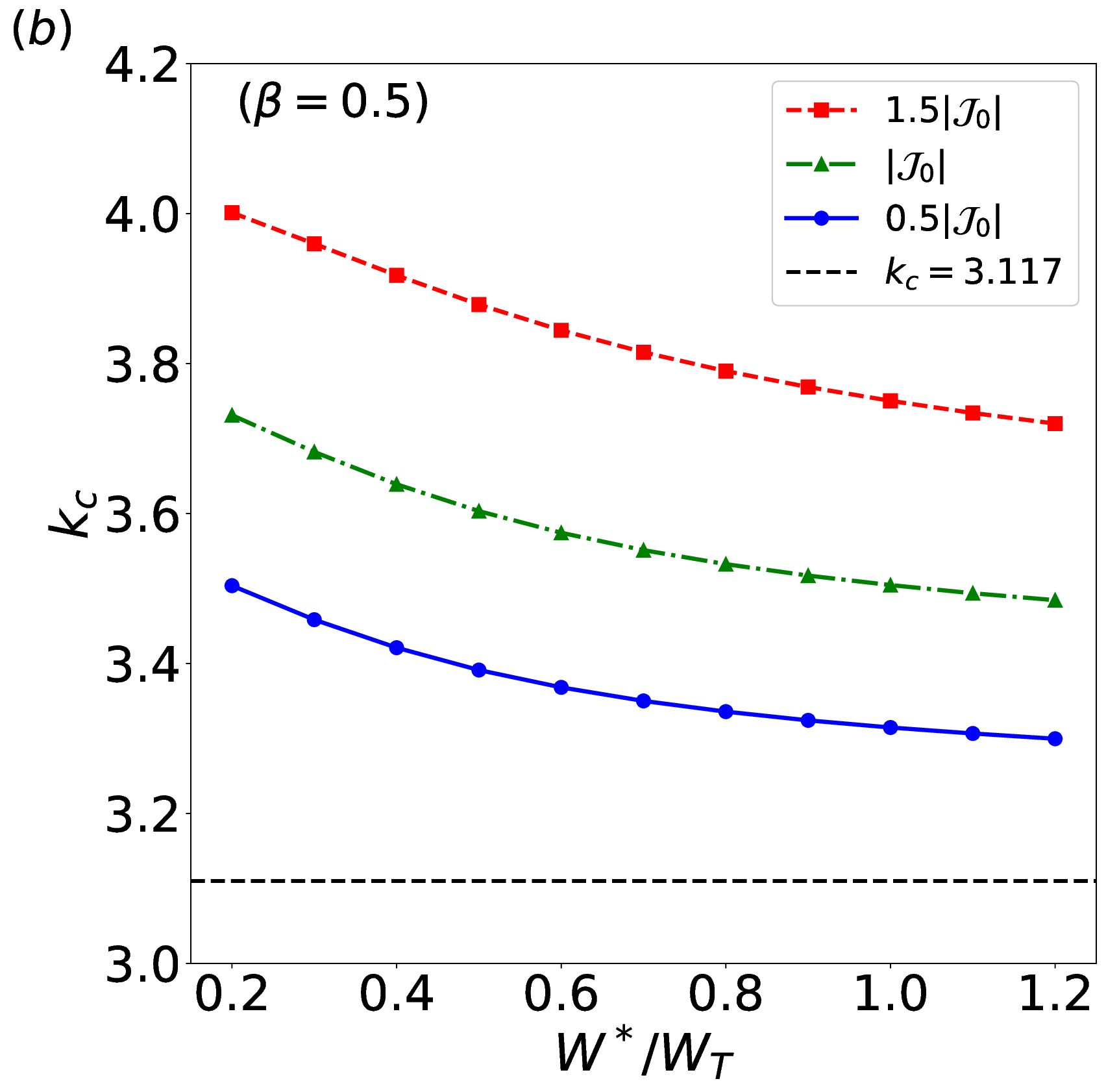}
  \label{fig:fig1}
\end{minipage}\hfill
\begin{minipage}[t]{0.24\textwidth}
\includegraphics[width=\linewidth]{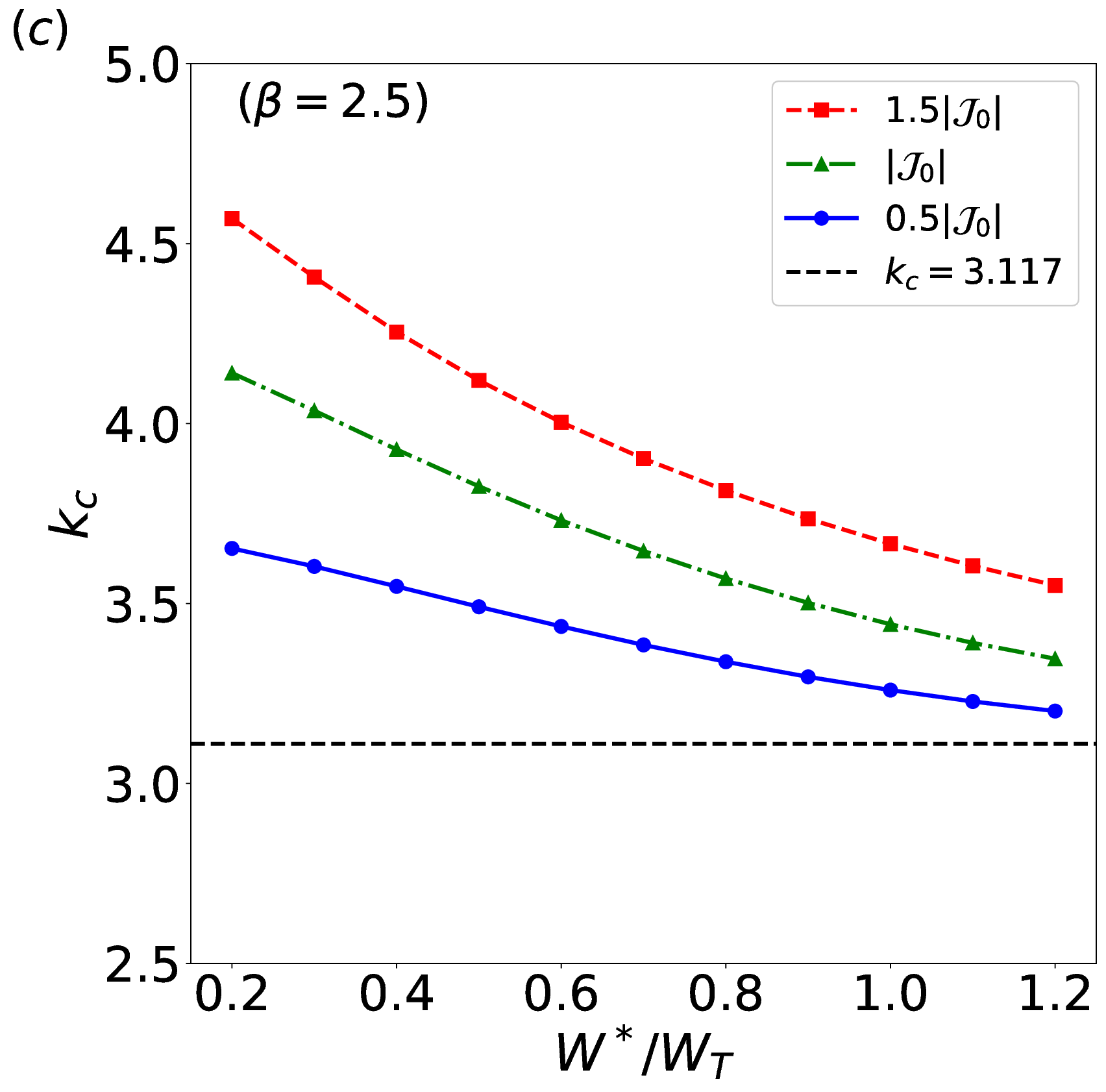}
  \label{fig:fig1}
\end{minipage}\hfill
\begin{minipage}[t]{0.24\textwidth}
\includegraphics[width=\linewidth]{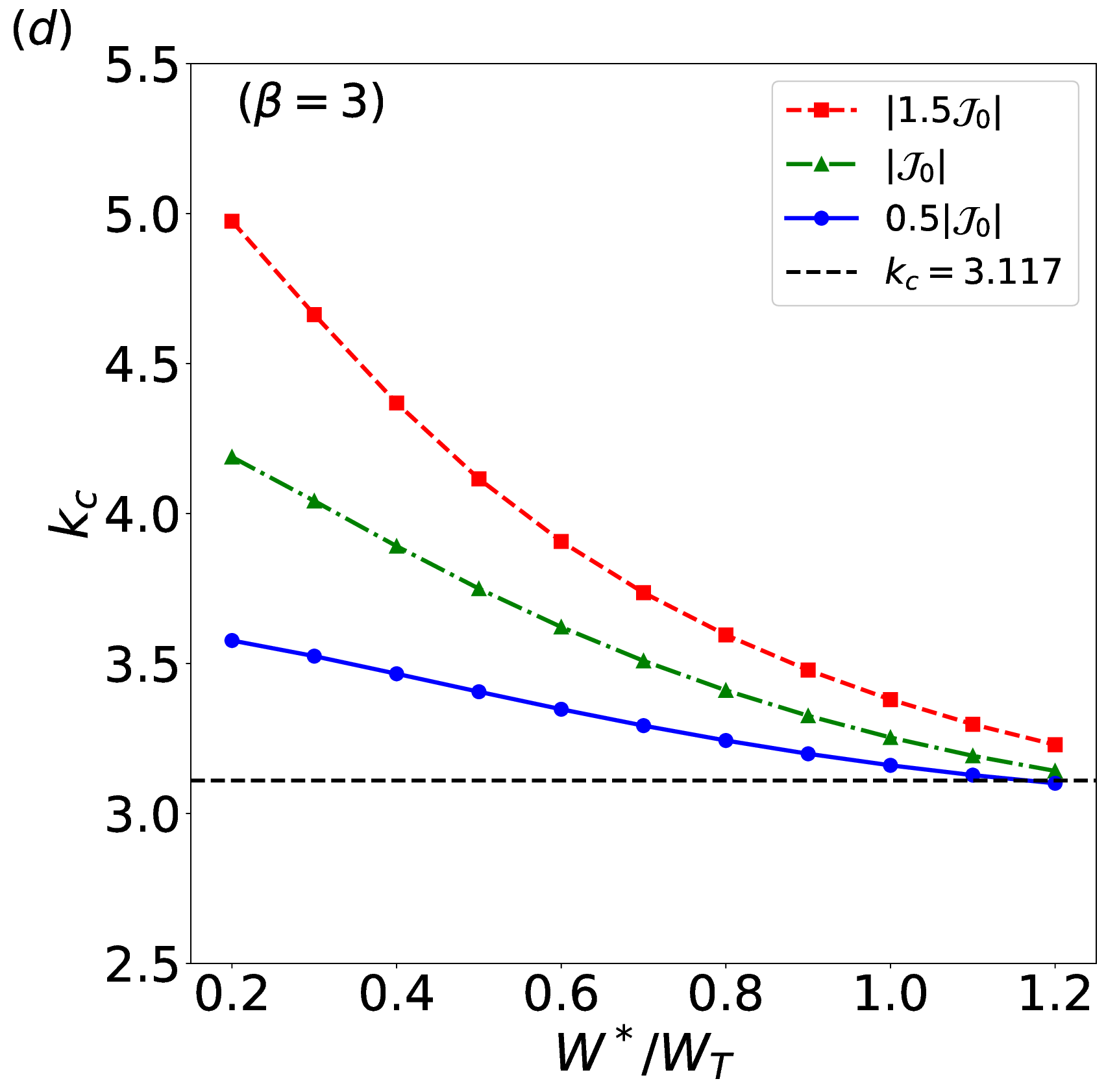}  
  \label{fig:fig1}
\end{minipage}
\vspace{-5mm}
\caption{Same as figure \ref{fig:all_figures-flux}, but for the critical wave number.}
    \label{fig:all_figures-flux-kc}
\end{figure}

For heavy particles, e.g. figures \ref{fig:all_figures-flux}(a) and \ref{fig:all_figures-flux}(b), increasing the particle inlet flux is always stabilizing, but the effect is more pronounced at low inlet velocities. Additional simulations were conducted for heavy particles, exploring higher values of both particle flux and inlet velocity, within the validity range of the present model. In all cases, the critical Rayleigh number remained above the RB threshold, providing strong evidence that it is not possible to destabilize the RB model by adding heavy particles. Furthermore, figures \ref{fig:all_figures-flux-kc}(a)-(d) show that the wave number dependence on both particle flux and inlet velocity is relatively weaker. This means that the size of the convective cells at the onset of convection are not significantly different from their single-phase RB counterparts. Such a difference becomes more pronounced, however, for very slow-rising bubbles at high volumetric inlet rates, as illustrated by the red line in figure \ref{fig:all_figures-flux-kc}(d).

\begin{figure}
    \centering
    % First Row
    \begin{minipage}[t]{0.32\textwidth}
        \centering
        {\footnotesize (a) $\beta=0.5\ ,\  W^*/W_T =  0.5$}\\
        \includegraphics[width=\textwidth]{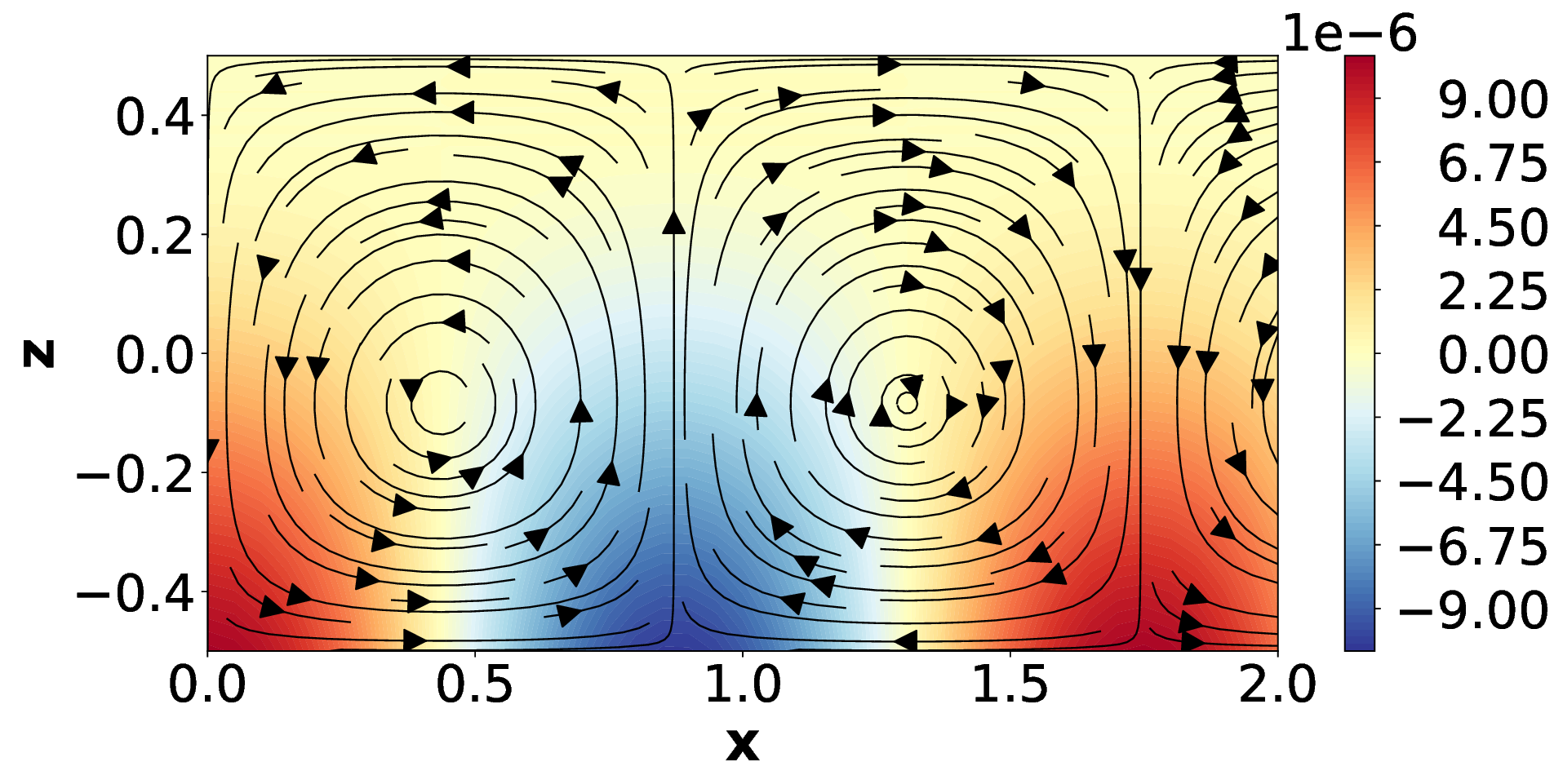}
        %\caption*{(a) $W^*/W_T =  0.5$}
    \end{minipage}
    \hfill
    \begin{minipage}[t]{0.32\textwidth}
        \centering
        {\footnotesize (b) $\beta=0.5\ ,\  W^*/W_T =  0.8$}\\
        \includegraphics[width=\textwidth]{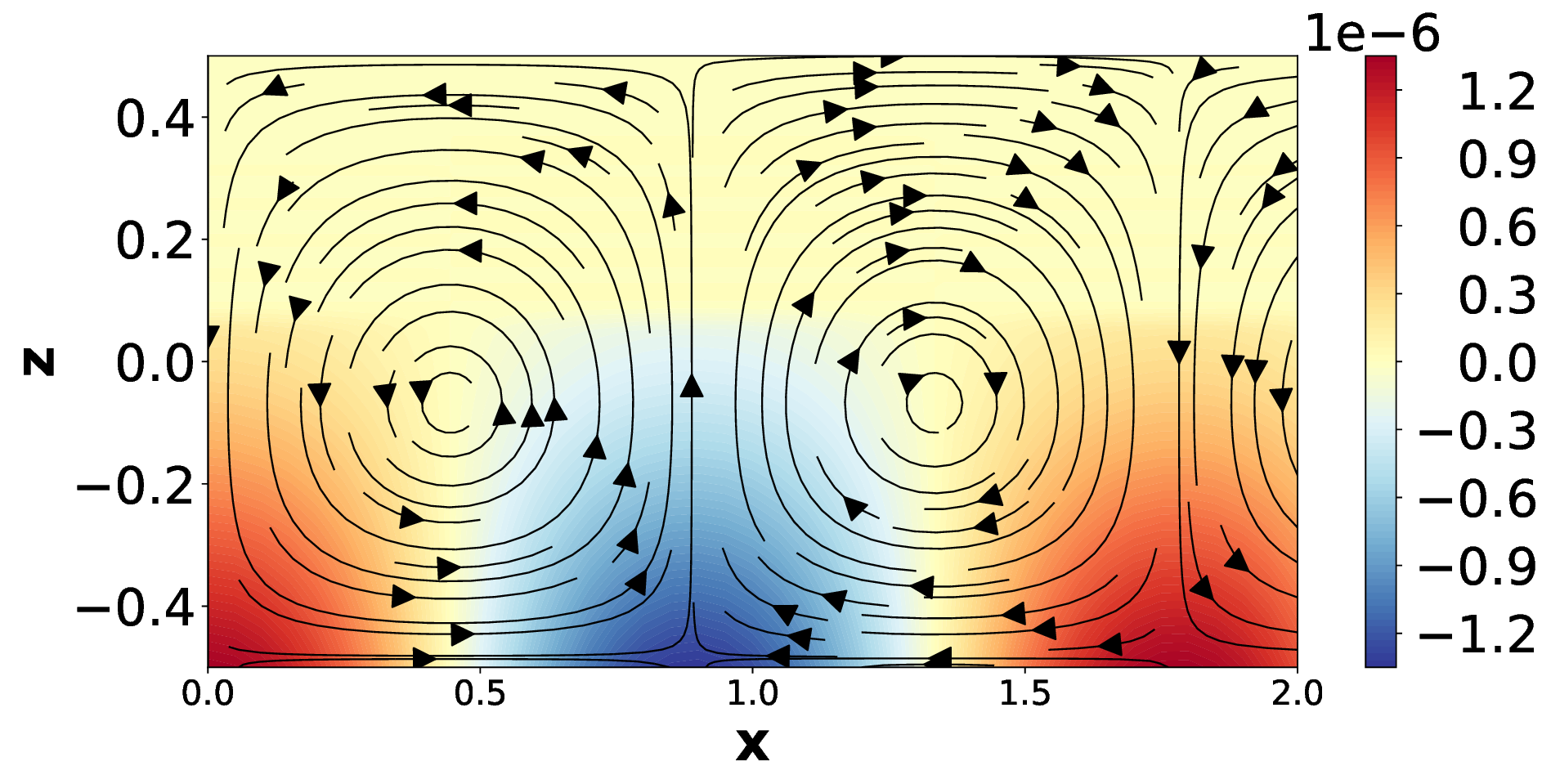}
        %\caption*{(b) $W^*/W_T =  0.8$}
    \end{minipage}
    \hfill
    \begin{minipage}[t]{0.32\textwidth}
        \centering
        {\footnotesize (c) $\beta=0.5\ ,\  W^*/W_T =  1.2$}\\
        \includegraphics[width=\textwidth]{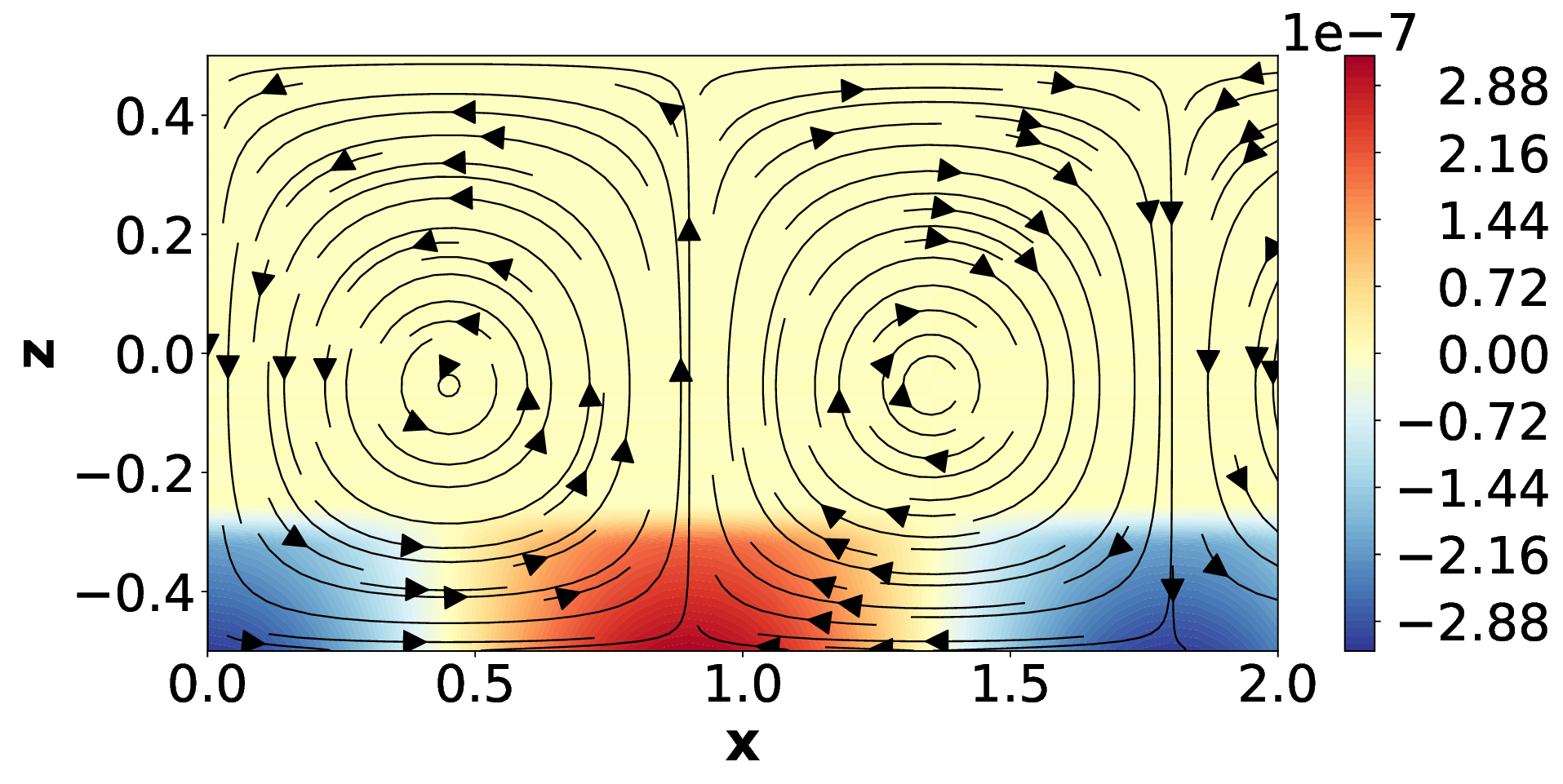}
        %\caption*{(c) $W^*/W_T =  1.2$}
    \end{minipage}
    
    %\vspace{0.1cm} % space between rows
    
    % Second Row
    \begin{minipage}[t]{0.32\textwidth}
        \centering
        {\footnotesize (d) $\beta=3\ ,\  W^*/W_T =  0.5$}\\
        \includegraphics[width=\textwidth]{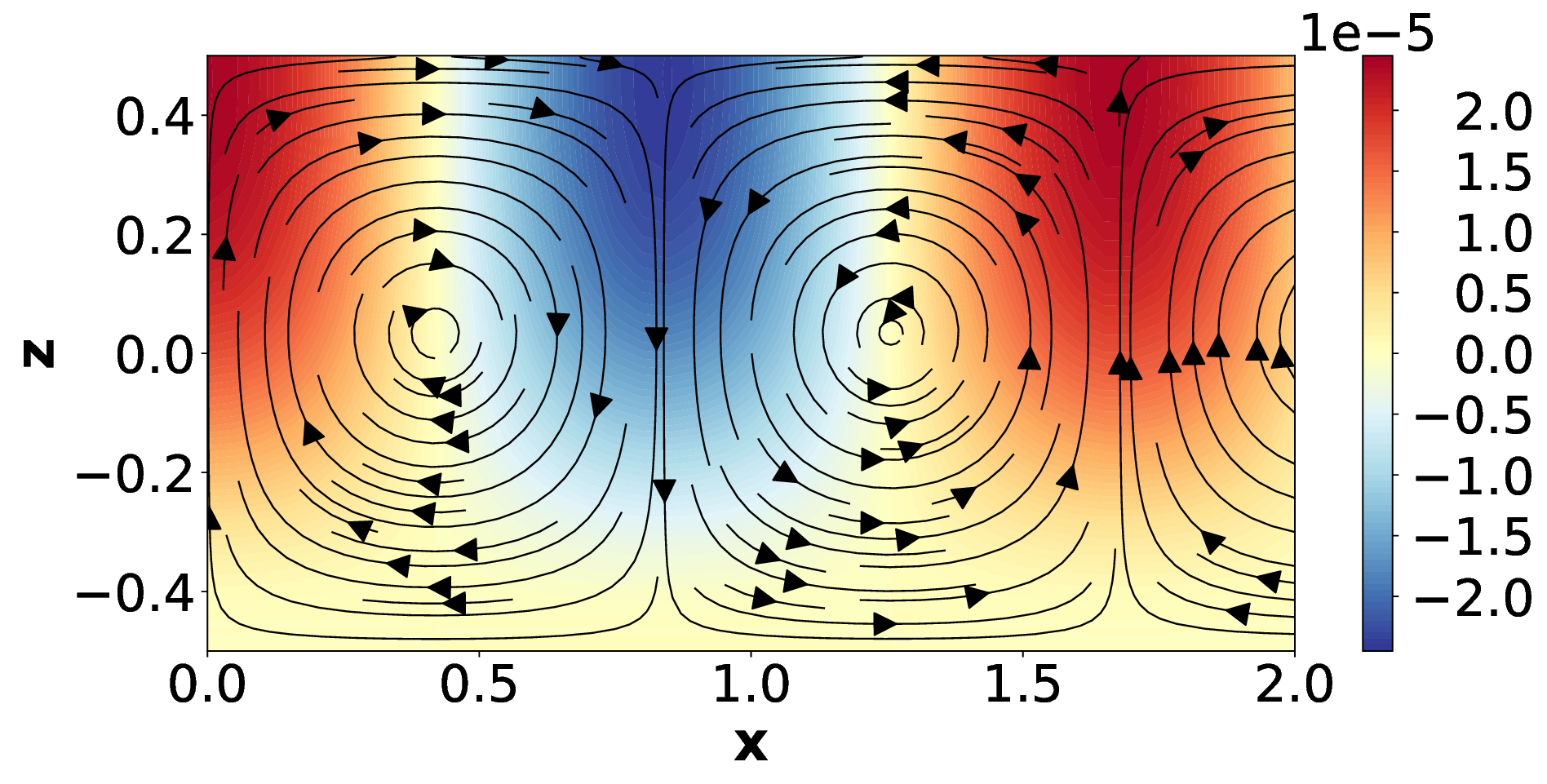}
        %\caption*{(d) $W^*/W_T =  0.5$}
    \end{minipage}
    \hfill
    \begin{minipage}[t]{0.32\textwidth}
        \centering
        {\footnotesize (e) $\beta=3\ ,\  W^*/W_T =  0.8$}\\
        \includegraphics[width=\textwidth]{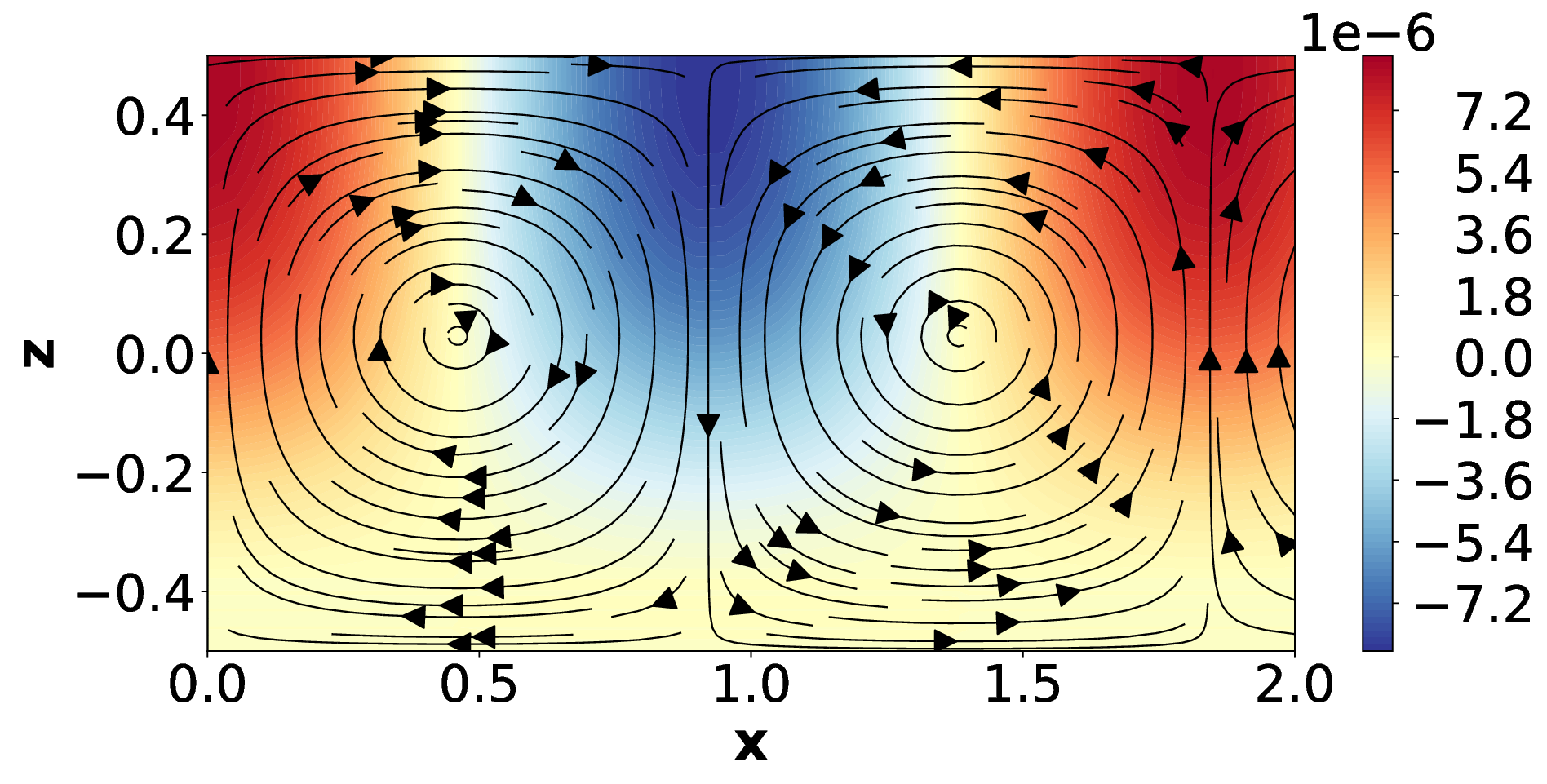}
        %\caption*{(e) $W^*/W_T =  0.8$}
    \end{minipage}
    \hfill
    \begin{minipage}[t]{0.32\textwidth}
        \centering
        {\footnotesize (f) $\beta=3\ ,\  W^*/W_T = 1.2$}\\
        \includegraphics[width=\textwidth]{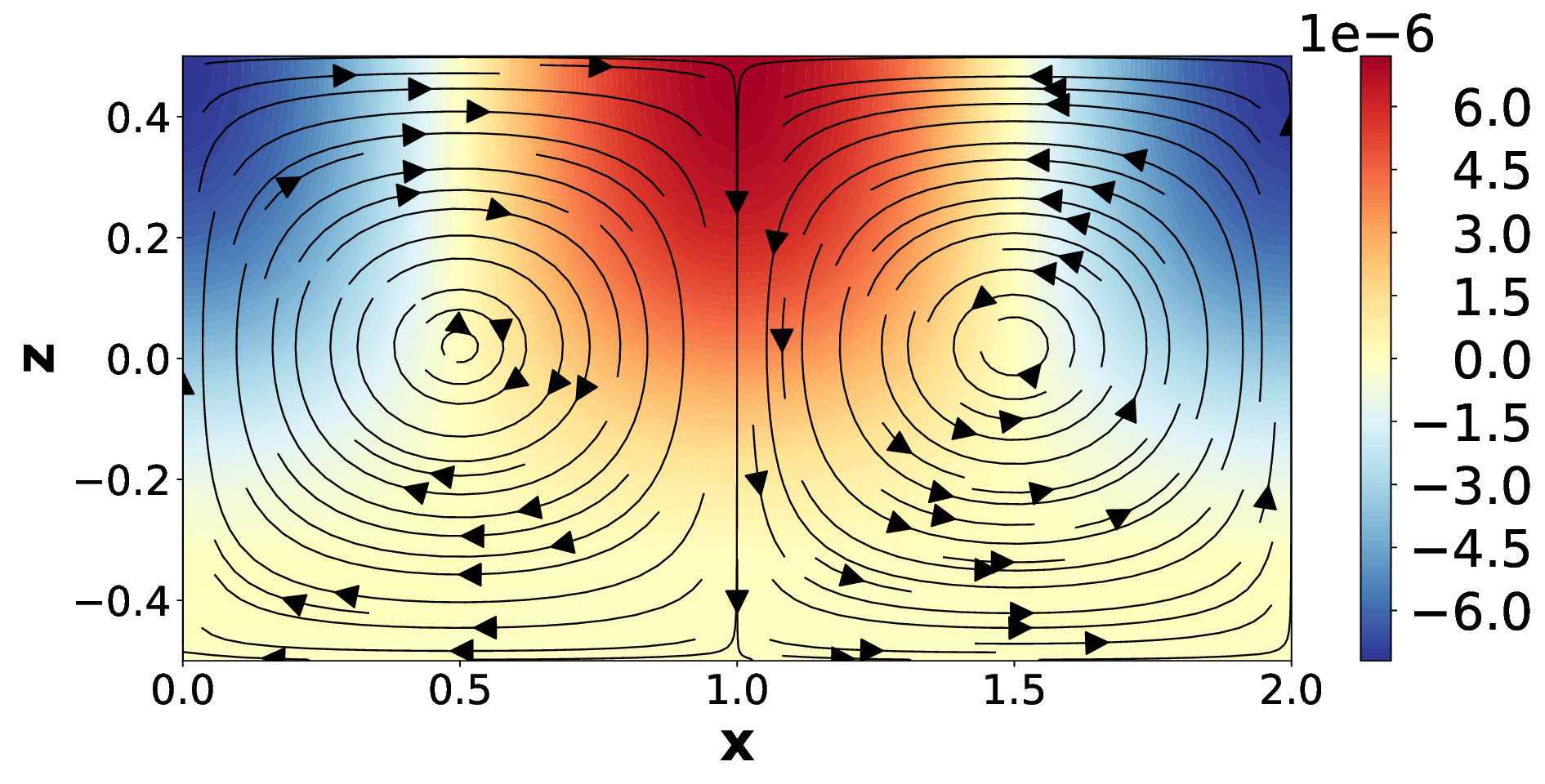}
        %\caption*{(f) $W^*/W_T =  1.2$}
    \end{minipage}\\[1ex]
     \caption{Streamlines of fluid velocity field and colormap of the particle volume fraction at the onset of convection for $\beta = 0.5$ (a-b-c) and  $\beta = 3$ (d-e-f) and increasing inlet velocities from left to right. For a better comparison, the particle volume fraction $\alpha'$ is normalized with respect to the base volume fraction $\alpha_0$ for $\beta=3$ and $W^*=W_T$ particles. \textcolor{black}{The particle inlet flux is $\mathcal{J} = \mathcal{J}_0$}}
     \label{fig:all_figures-iso}
\end{figure}

Streamlines and particle concentration at the onset of convection are plotted in figure \ref{fig:all_figures-iso}, through the eigenvectors of the perturbed fluid velocity $\bm{U'}$ and particle distribution $\alpha'$, for both heavy and light particles. Heavy particles (a-c) accumulate near the bottom. For sub-terminal (super-terminal) inlet velocities, a higher particle volume fraction is observed in the downwelling (upwelling) plumes. The particular case where $W^*=W_T$ is not shown for conciseness as it is already known that it possess a uniform particle distribution \citep{razaPF2024}. The opposite behavior is observed for light particles (d-f). They accumulate near the top, with upwelling (downwelling) plumes favoring higher concentrations for sub-terminal (super-terminal) velocities. Finally, when comparing figures \ref{fig:all_figures-iso}(c) and \ref{fig:all_figures-iso}(f), it can be observed that heavy particles ($\beta=0.5$) tend to accumulate closer to the walls than light ones ($\beta=3$) at super-terminal velocities ($W^*/W_T=1.2$).

\subsection{Energy budget analysis}

\begin{figure}
    \centering
    \includegraphics[width=0.45\linewidth]{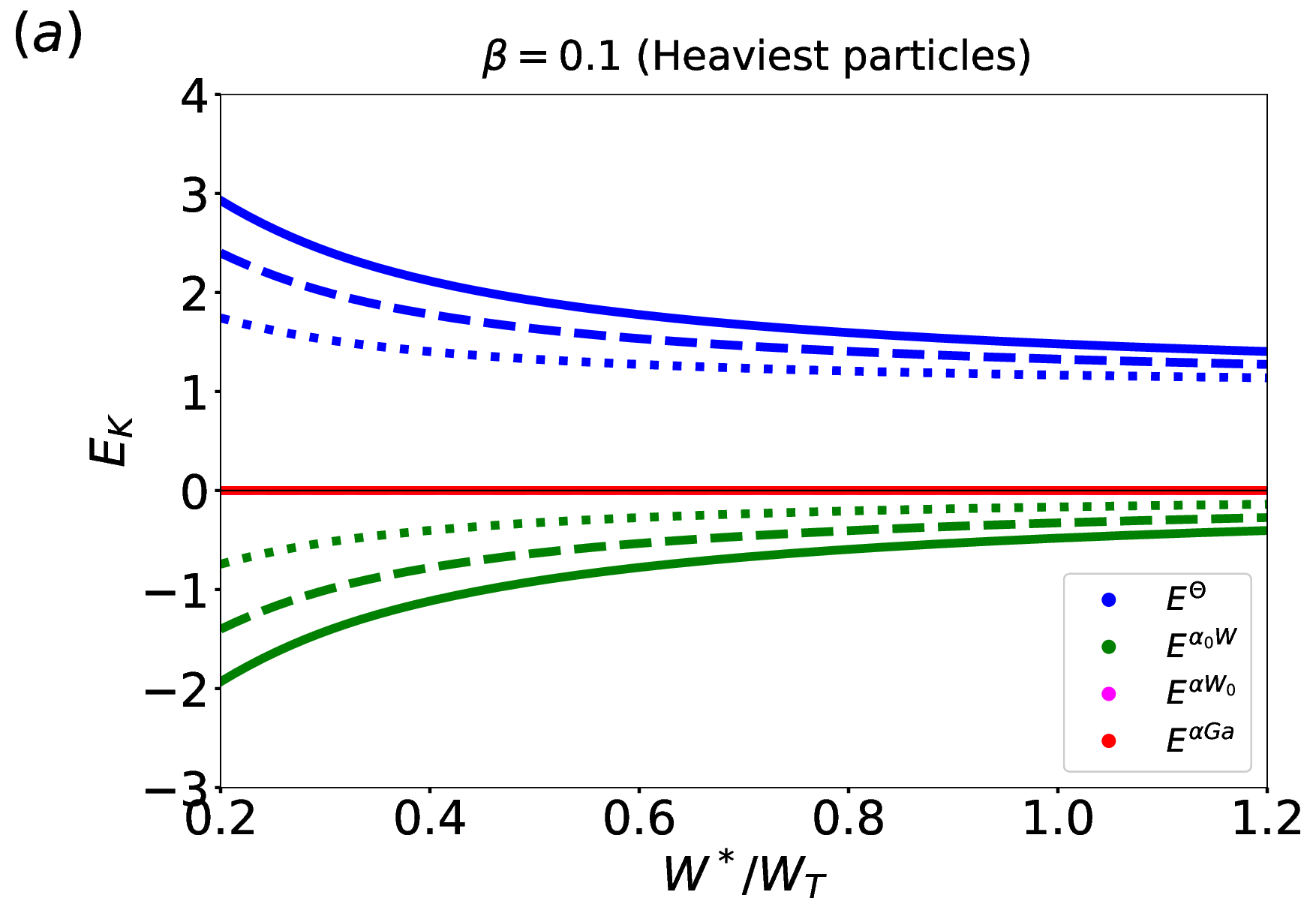} 
    \includegraphics[width=0.45\linewidth]{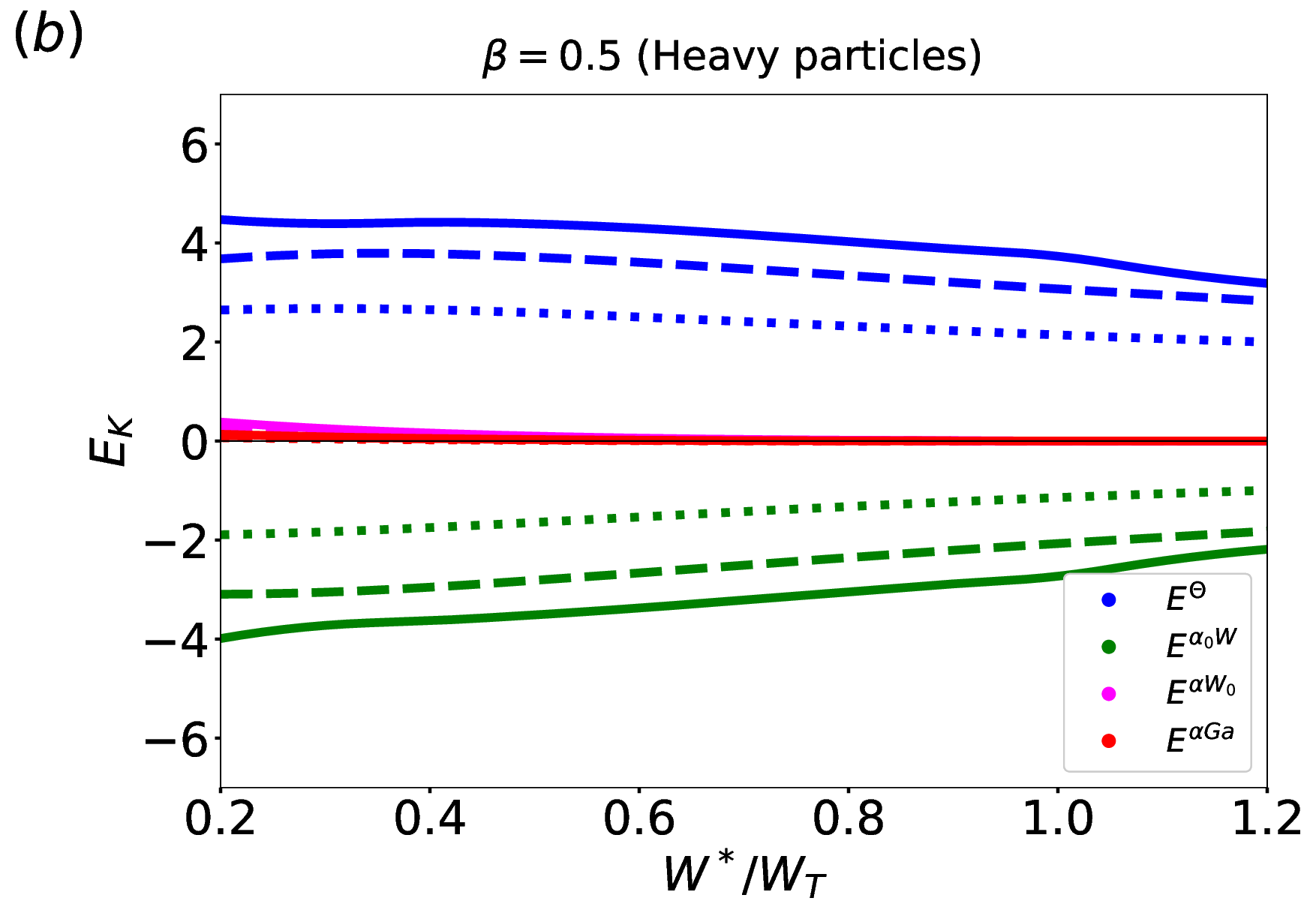} \\[1ex]
    \includegraphics[width=0.45\linewidth]{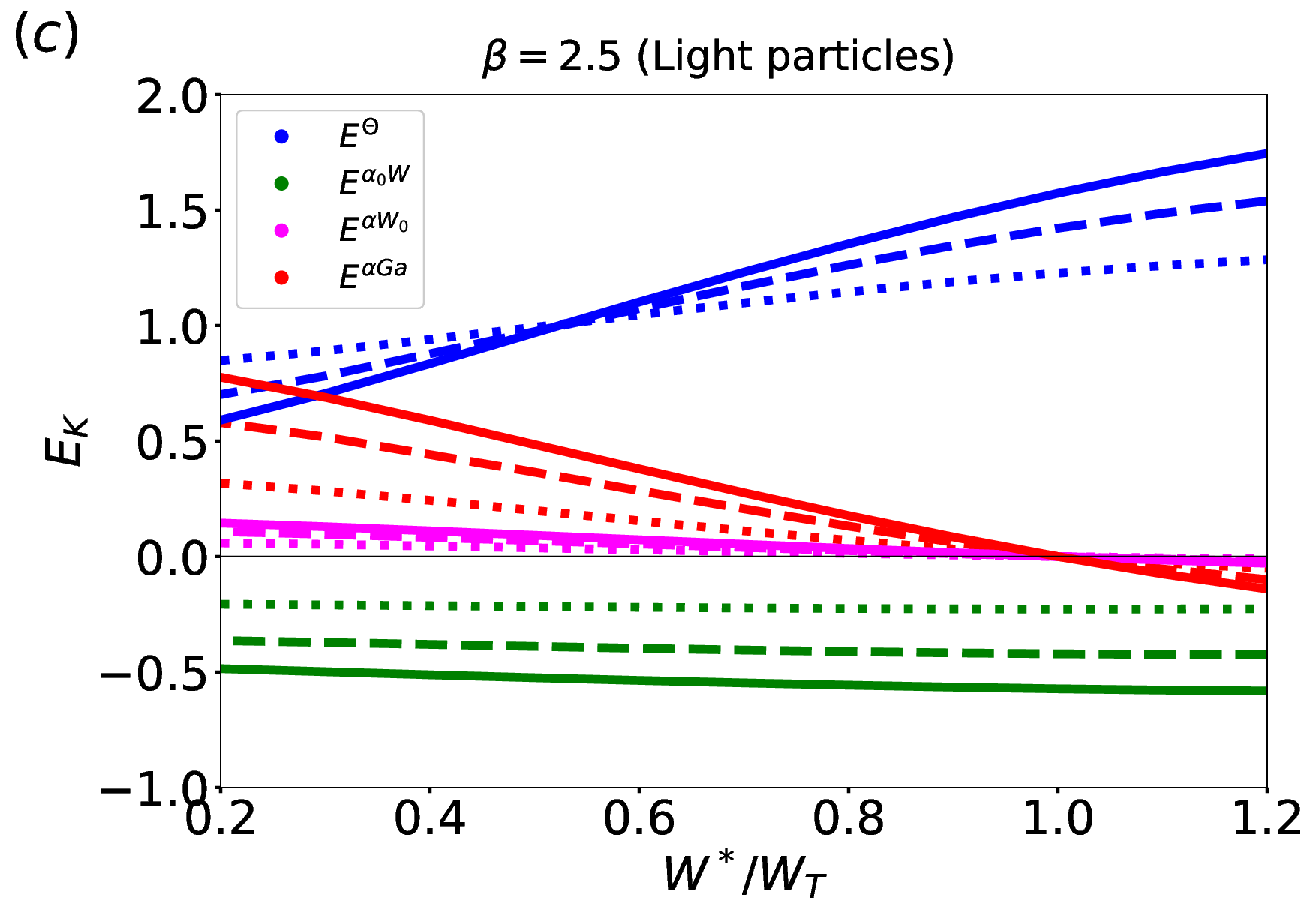} 
    \includegraphics[width=0.45\linewidth]{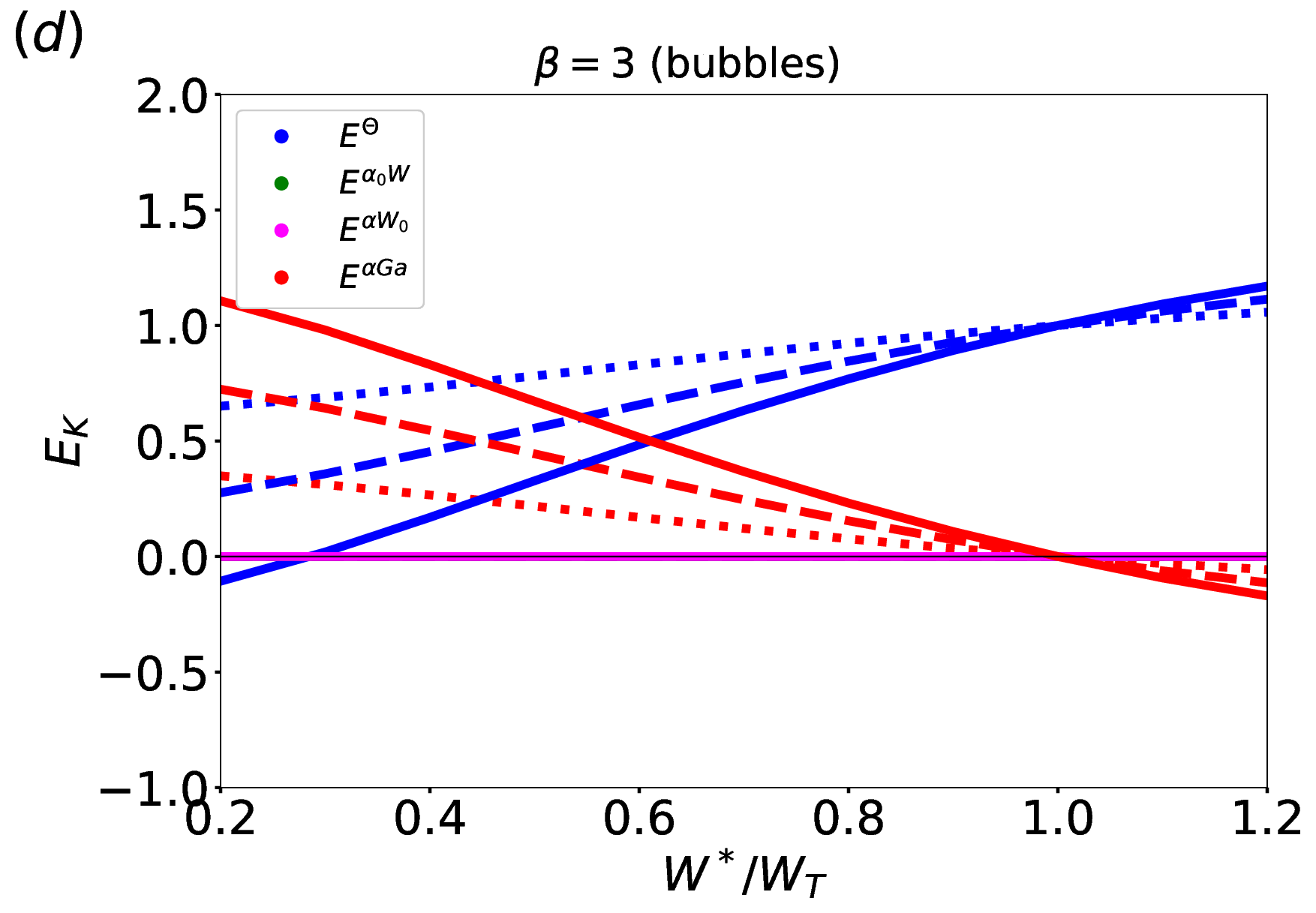} 
    \caption{Components of the kinetic energy budget at the neutral stability condition versus the particle inlet velocity: thermal buoyancy injection power $E^{\Theta}$, viscous dissipation rate $E^{V}$, particle feedback's due to the base particle concentration, base particle velocity and particle buoyancy $E^{\alpha_0 W}$, $E^{\alpha W_0}$,$E^{\alpha  Ga}$. Results obtained for different particle inlet fluxes: $0.5\mathcal{J}_0$ (dotted lines), $\mathcal{J}_0$ (dashed lines), $1.5\mathcal{J}_0$ (solid lines).}
    \label{fig:energy-budget}
\end{figure}

An energy budget analysis is performed in order to gain further insights into the physical mechanisms triggering instabilities in the pRB system. In order to derive the evolution equation for the perturbation kinetic energy, we take the product of the momentum equation (\ref{eq35}) and the complex conjugate of the vertical fluid velocity, $\bar{U}_z^n$, where the overbar denotes complex conjugate. By integrating the resulting equation over the entire domain, the following relationship for the spatially averaged perturbation kinetic energy rate is obtained: 
\begin{equation}
    \lambda\ e^K= e^{\Theta} + e^{V} + e^{\alpha_0W} + e^{\alpha W_0}+e^{\alpha Ga}
\end{equation}
with each term in this equation being defined as:
\begin{align}
         e^K &= \left(1 - \tfrac{\alpha_0}{2} (\beta -1) \right) \scaleobj{.7}{\int_{-1/2}^{1/2}} \mathrm{Re}\left[(D^2 -k^2) (U_z^n)  \bar{U}_z^n \right]dZ\notag\\
         &{\quad +\tfrac{(\beta -1)}{2}\scaleobj{.7}{\int_{-1/2}^{1/2}} D\alpha_0\mathrm{Re}\left[D (U_z^n)  \bar{U}_z^n \right]dZ},
         \label{eq:eK}\\
      e^{\Theta} &=  Pr Ra \scaleobj{.7}{\int_{-1/2}^{1/2}} \mathrm{Re}\left[(\Theta^n k^2) \bar{U}_z^n\right]dZ,\label{eq:eTheta}\\
        e^{V} &= -Pr \scaleobj{.7}{\int_{-1/2}^{1/2}} \mathrm{Re}\left[(D^2 -k^2)^2 (U_z^n)  \bar{U}_z^n \right]dZ,\label{eq:eV}\\
      e^{\alpha_0 W} &= \tfrac{6\, \mathrm{Pr} (3 - \beta)}{\Phi^2}\bigg[
\scaleobj{.7}{\int_{-1/2}^{1/2}} \alpha_0 \, \mathrm{Re} \Big[ (D^2 - k^2) (U_z^n)  \bar{U}_z^n + (\dot{\iota} k D W_x^n + k^2 W_z^n)  \bar{U}_z^n \Big]dZ \notag \\
&{\quad +\scaleobj{.7}{\int_{-1/2}^{1/2}} D\alpha_0 \mathrm{Re} \Big[ (\dot{\iota} k W_x^n)\bar{U}_z^n + D (U_z^n)  \bar{U}_z^n \Big]} \ dZ\bigg],\label{eq:ea0W}\\
     e^{\alpha W_0} &=  \tfrac{6 Pr (3 -\beta)}{\Phi^2} \scaleobj{.7}{\int_{-1/2}^{1/2}} W_0 \mathrm{Re}\left[ (\alpha^n k^2)  \bar{U}_z^n 
 \right]dZ,\label{eq:eaW0}\\
        e^{\alpha Ga} &=  {\tfrac{(\beta -1) {Ga Pr^2}}{2}}\scaleobj{.7}{\int_{-1/2}^{1/2}} \mathrm{Re}\left[ (\alpha^n k^2)  \bar{U}_z^n 
 \right]dZ.\label{eq:eGa}
\end{align}
where $e^{\Theta}$ is the power of the thermal buoyancy force, $e^{V}$ indicates the rate of viscous energy dissipation. Furthermore, $e^{\alpha_0 W}$, $e^{\alpha W_0}$ and $e^{\alpha  Ga}$ collectively represent the particle feedback, i.e. the parts due to the base particle concentration, base particle velocity and particle buoyancy, respectively. The first two terms represent the drag force, while the latter represents the Archimedes force. The superscript $\alpha$ denotes the terms arising from the non-homogeneous distribution of particles, which are absent if $W^*=W_T$ as in \cite{prakhar2021linear, razaPF2024}. By normalizing these contributions using the absolute value of the dissipation rate, we obtain at neutral condition $(\lambda = 0)$:
\begin{equation}
   E^{\Theta}  + E^{\alpha_0W} + E^{\alpha W_0}+ E^{\alpha Ga} = 1.
   \label{kineticbudget}
\end{equation}
Figure~\ref{fig:energy-budget} shows the aforementioned normalized kinetic energy rates versus the particle inlet velocity for different particle fluxes \( \mathcal{J} \) and density ratios \( \beta \). Positive (negative) contributions indicate destabilization (stabilization) of the steady-state.
Heavy-particle cases (figures \ref{fig:energy-budget}(a, b)) show a \Saad{competition} between the destabilizing thermal buoyancy and the stabilizing particle drag feedback. Hence, heavy particle action is purely dissipative as their friction opposes the thermal buoyancy. However, the energy injection due to particle buoyancy  \( E^{\alpha Ga} \) is always negligible in the present study. This is at odds with the light particle cases (figures ~\ref{fig:energy-budget}(c, d)) where the particle buoyancy term turns out to be important. Here \( E^{\alpha Ga} \) is destabilizing when the injection velocity is smaller than the terminal one, while the opposite is true otherwise. This highlights the importance of the direction of particle motion, namely ascending or descending, as well as the sign of their acceleration, which determines whether the particle concentration near the injection wall is diluted or intensified. 
We note that in the special case of bubbles (figure ~\ref{fig:energy-budget}(d)), where the drag terms \( E^{\alpha_0 W} \) and \( E^{\alpha W_0} \) vanish by definition according to equations (\ref{eq:ea0W}–\ref{eq:eaW0}), the competition is solely between the thermal and particle buoyancies.
\section{Concluding remarks}
The present investigation highlighted how the particle inlet velocity influences the linear stability of the pRB system. Increasing the inlet velocity while maintaining a constant particle flux destabilizes the system for heavy particles but progressively stabilizes it for light particles. While the general features of particle accumulation persist across the range of inlet velocities, the spatial localization of accumulation shifts between upwelling and downwelling regions as the injection speed transitions from sub-terminal to super-terminal values. When the injection velocity matches the terminal velocity, the linearized system dynamics does not support the accumulation of particles. \textcolor{black}{In that case particle accumulation could arise only through nonlinear interactions
as shown by \citet{srinivas2025}.} In this study the thermal coupling was deliberately minimized in order to isolate the effects of the mechanical coupling introduced by the injection velocity. This was achieved by maintaining $E \ll 1$ and setting the particle inlet temperature equal to the inlet-wall temperature. Nevertheless, particle thermal inertia and injection temperature offer alternative mechanisms for influencing the flow onset within this model. 
%and will be explored using the same stability analysis framework.
\textcolor{black}{Preliminary investigations in this direction \citep{razajapan2025} indicate that increasing $E$ consistently enhances system stability when particles are heavier than the fluid, regardless of the volumetric particle flux, injection velocity, or injection temperature. In contrast, for particles lighter than the fluid, the effects are more complex: the influence of $E$ can be either stabilizing or destabilizing, depending on the inlet velocity and particulate volumetric flux. This suggest that the effects of momentum coupling dominate over the thermal ones. However, we shall note that the parameter $E$ depends on the relative density between the particles and the fluid, in fact $E=(c_{Pp}/c_P)(3-\beta)/(2\beta)$ implying that when the particles are very light, the thermal coupling decreases and eventually becomes negligible in the bubble limit. The physical mechanisms underlying the behavior of the thermal coupling remain to be fully understood and will be the focus of future work.\\
Finally, an essential direction for future investigation of this problem is the impact of particle size. This can be addressed only in a very limited way within the framework of an Eulerian model, such as the one considered here—for example, by including Faxén corrections in the particle equations. However, it can be more effectively approached through numerical studies with resolved particles, such as those recently proposed by \cite{ChenPropseretti2024}.}\\

\textbf{Declaration of interests.} The authors report no conflict of interest.

%\bibliographystyle{jfm}
%\bibliography{PRBbiblio}

\end{document}